\documentclass[journal]{IEEEtran}
\ifCLASSINFOpdf
\else
\fi

\usepackage{xcolor}

\usepackage{framed} 

\colorlet{shadecolor}{yellow}
\usepackage[pdftex]{graphicx}
\graphicspath{{../pdf/}{../jpeg/}}
\DeclareGraphicsExtensions{.pdf,.jpeg,.png}

\usepackage[cmex10]{amsmath}
\usepackage{array}
\usepackage{mdwmath}
\usepackage{mdwtab}
\usepackage{eqparbox}
\usepackage{url}

\usepackage{amssymb}
\graphicspath{ {./images/} }
\usepackage{orcidlink}

\usepackage{breqn}
\usepackage{physics}

\usepackage{tikz} 

\usepackage{soul}   

\usepackage{enumitem}  

\usepackage{threeparttable}  

\usepackage{makecell} 

\newif\ifshowmods

\ifshowmods
                 
                \newcommand{\Ablue}[1]{\Ablue{#1}} 
                \newcommand{\Ablack}[1]{\textcolor{black}{#1}}

\else

                \newcommand{\Ablue}[1]{#1}
                \newcommand{\Ablack}[1]{#1}

\fi

\usepackage{amsmath}

\begin{document}

\title{






\Ablack{An AI-Based Supervisory Measurement Integrity Validation Layer for Cyber-Resilient  AC/DC Protection in  Inverter-Based Microgrids}

}

\author{
    Ahmad~Mohammad~Saber\orcidlink{0000-0003-3115-2384},~\IEEEmembership{Member,~IEEE,} 
    Ahmed~Saber~Refae\orcidlink{0000-0001-5147-6402}, 
    Davor~Svetinovic\orcidlink{0000-0002-3020-9556}, 
    Hatem~Zeineldin\orcidlink{0000-0003-1500-1260}, 
    Amr~Youssef\orcidlink{0000-0002-4284-8646},~\IEEEmembership{Senior Member,~IEEE,}
     Ehab~F.~El-Saadany\orcidlink{0000-0003-0172-0686},~\IEEEmembership{Fellow,~IEEE}, 
    and Deepa~Kundur\orcidlink{0000-0001-5999-1847},~\IEEEmembership{Fellow,~IEEE}

\thanks{\Ablack{Ahmad Mohammad Saber and Deepa Kundur are with the Department of Electrical and Computer Engineering, University of Toronto, Toronto, ON, Canada
(e-mails: \href{mailto:ahmad.m.saber@ieee.org}{ahmad.m.saber@ieee.org}, \href{mailto:dkundur@ece.utoronto.ca}{dkundur@ece.utoronto.ca}).}}
\thanks{\Ablack{Ahmed Saber Refae is with the
Electric Power Engineering Department, Cairo University, Giza, Egypt
(e-mail: a\_saber\_86@cu.edu.eg).
}}
\thanks{\Ablack{Davor Svetinovic, Hatem H. Zeineldin and Ehab F. El-Saadany are with the Department of Electrical Engineering, Khalifa University, Abu Dhabi, UAE (emails: \href{mailto:davor.svetinovic@ku.ac.ae}{davor.svetinovic@ku.ac.ae},  \href{mailto:hatem.zeineldin@ku.ac.ae}{ hatem.zeineldin@ku.ac.ae},  \href{mailto:ehab.elsadaany@ku.ac.ae}{ehab.elsadaany@ku.ac.ae}).}}
\thanks{\Ablack{Amr Youssef is with
the Concordia Institute for Information Systems Engineering (CIISE),
Concordia University, Montreal, QC, Canada
(e-mail: 
\href{mailto:youssef@ciise.concordia.ca}{youssef@ciise.concordia.ca}).}}
}

\markboth{
}%
{Shell \MakeLowercase{\emph{et al.}}: Bare Demo of IEEEtran.cls for IEEE Journals}

\maketitle

\begin{abstract}

\Ablack{Line current differential relays (LCDRs) are measurement-driven relays that rely on time-synchronized multi-phase current waveforms to infer internal faults in AC and DC power networks. In inverter-based microgrids, however, the increasing reliance on digitally communicated measurements exposes LCDRs to false-data injection attacks (FDIAs), in which adversaries manipulate remote \textit{measurement} streams to create protection-triggering yet physically inconsistent current trajectories. This paper addresses this emerging \textit{measurement integrity} problem by introducing a measurement integrity validation scheme that operates as a supervisory instrumentation layer for modern LCDRs.
The proposed scheme interprets short windows of synchronized instantaneous current measurements recorded during relay operation and assesses their physical consistency to distinguish genuine fault-induced trajectories from cyber-manipulated measurement streams. A recurrent neural network is trained offline using only relay-available current measurements and exploits the temporal structure of differential current waveforms, which remains informative in inverter-dominated systems where current magnitude is no longer a reliable observable. The method requires no additional sensors, auxiliary protection elements, or prior knowledge of network topology, and is applicable to both AC and DC LCDRs without structural modification.
The proposed measurement validation scheme is evaluated on an islanded inverter-based microgrid under a comprehensive set of fault and FDIA scenarios, demonstrating high detection accuracy while preserving relay dependability. Hardware-in-the-loop validation using an OPAL-RT real-time simulator confirms that the scheme satisfies protection timing constraints and can operate in real time under realistic operating conditions.}

\end{abstract}

\begin{IEEEkeywords}

Cyber-physical security,
false-data-injection attacks,   
inverter-based microgrids,
line current differential relays,
protection.



\end{IEEEkeywords}

\IEEEpeerreviewmaketitle

\section{Introduction}
\label{section:Introduction}

\Ablue{\IEEEPARstart{T}{he} increasing integration of information and communication technologies (ICTs) into 
modern power systems has expanded the cyber attack surface of measurement-driven 
instrumentation and control. False-data injection attacks (FDIAs), in which an 
adversary deliberately manipulates measurement streams, directly compromise the 
integrity of the measurement process itself, potentially inducing incorrect inference 
about the underlying physical state even when the physical system remains intact 
\cite{Reiter_FDIA}. The real-world consequences of such attacks were highlighted by 
the 2015 cyber incident on the Ukrainian power grid, which disrupted electricity 
service for hundreds of thousands of customers. As microgrids are increasingly deployed 
to enhance resilience for critical infrastructures, such as military installations 
\cite{Siemens_MG_Military}, their reliance on synchronized digital measurements makes 
their protection components especially attractive targets for cyber adversaries 
\cite{MG_protection, Appl_energy_MG_review_paper2}.}

\Ablack{Line current differential relays (LCDRs) are among the most measurement-intensive protection instruments in both transmission systems and microgrids. They are widely adopted due to their high speed, sensitivity, and selectivity compared to overcurrent- and distance-based relays \cite{MG_protection,CyberResilient_TIM_DistanceRelay}. In inverter-based microgrids, LCDRs are particularly attractive because fast fault isolation is critical to maintaining stability in systems characterized by low inertia and limited fault current contribution \cite{Hatem_Diff_MIIMs}. Modern LCDRs operate by comparing time-synchronized local and remote current measurements, relying on Kirchhoff’s current law to infer the presence of internal faults \cite{howLCDRworks}. This dependence on remotely communicated measurements, however, makes LCDRs especially vulnerable to FDIAs that manipulate synchronized current data streams \cite{Reiter_FDIA}.}

\Ablack{From a measurement perspective, LCDRs can be broadly classified into phasor-based schemes, which exchange root-mean-square (r.m.s.) values or phasors estimated over one or more cycles, and sampled-value-based schemes, which operate directly on time-synchronized instantaneous current waveforms. They may further be categorized by operating principle into AC and DC LCDRs, and by application into transmission-level and microgrid-oriented relays. These distinctions are not merely taxonomic: they fundamentally affect how measurements should be interpreted. In inverter-based microgrids, fault currents are actively limited by converter controls, causing fault-induced current waveforms to lack the pronounced magnitude changes assumed by phasor-based and steady-state measurement techniques. As a result, instantaneous time-domain measurements and their temporal structure become the primary observable for reliable fault characterization.}

\Ablack{This shift in measurement modality introduces a new challenge. While sampled-value-based LCDRs improve sensitivity in inverter-dominated systems, they also expose raw time-domain measurement streams to cyber manipulation. Under FDIA conditions, adversaries can craft measurement trajectories that satisfy conventional relay operating criteria while remaining physically inconsistent with genuine fault-induced dynamics. Distinguishing between physically plausible fault-induced current trajectories and adversarially manipulated measurement streams thus becomes a nontrivial measurement interpretation problem, rather than a purely protection-logic problem.}

\Ablue{Moreover, recent literature has demonstrated the increased sophistication of FDIA strategies in power systems. New strategies include dynamic \cite{lu2022differential} and stealthy \cite{lu2021constrained} FDIA strategies based on differential evolution. This increased sophistication in FDIA strategies, whereby modern FDIAs are not limited to simple manipulations but can be systematically designed to mimic physically plausible behavior, underscores the need for the development of accurate defense mechanisms to be detect FDIAs against individual critical components in the smart grid such as LCDRs.}

\Ablack{Recent advances in artificial intelligence (AI) and machine learning offer new tools for addressing such challenges in measurement and instrumentation. Unlike classical threshold-based or feature-engineered approaches, AI models can learn the temporal consistency and multivariate dependencies inherent in synchronized measurement trajectories. This capability has motivated the use of deep learning for interpreting complex measurement data in cyber–physical systems, including FDIA detection in power transformers \cite{saber2025model,ali2023reliable}, transmission line protection \cite{mypaper_TII}, wide-area damping controllers \cite{zadsar2023preventing}, dynamic state estimation \cite{riahinia2024adaptive}, and overcurrent relays \cite{pola2023cyber}. These works demonstrate the growing role of AI as a measurement interpretation layer that complements conventional instrumentation pipelines. Importantly, they also highlight that effective AI-based measurement validation must be tailored to the physical dynamics and sensing modalities of each specific component.}

\Ablack{The cybersecurity of LCDRs has therefore attracted increasing attention in recent years \cite{Remedial_pilot_main_protection,mypaper_TII,Kundur_DL,Ahmad_smartgrid,Cyber_Resilient_Protection,Amir4,ResMVDC}. Early efforts primarily focused on DC LCDRs, proposing model- and threshold-based detection schemes that rely on DC-specific features or auxiliary hardware \cite{Cyber_Resilient_Protection,Amir4,ResMVDC}. While effective in limited settings, these approaches are not directly applicable to AC LCDRs and often require additional components that may be impractical in real systems. Other studies addressed cyberattacks on AC LCDRs in transmission networks powered by synchronous generators \cite{Remedial_pilot_main_protection,mypaper_TII,Kundur_DL,Ahmad_smartgrid}. In such systems, large fault current magnitudes provide strong measurement cues that simplify attack detection. However, these assumptions do not hold in inverter-based microgrids, where fault currents are inherently limited \cite{Saleh,Pirani2023}, rendering magnitude-based and phasor-based detection schemes ineffective.}

\Ablack{This paper addresses this gap by introducing a data-driven measurement integrity validation scheme (\Ablack{MIVS}) for LCDRs in islanded inverter-based microgrids. The proposed \Ablack{MIVS} interprets short windows of synchronized multi-phase current waveforms to assess their physical consistency and distinguish genuine fault-induced measurements from cyber-manipulated ones. Rather than modifying protection logic, the \Ablack{MIVS} operates as a supervisory measurement validation layer that confirms the plausibility of the measured physical phenomenon before tripping is permitted. The framework requires no additional sensors, auxiliary hardware, or system-specific features, and is applicable to both AC and DC LCDRs without structural modification. By exploiting the temporal structure of instantaneous current measurements using a recurrent neural network (RNN), the proposed \Ablack{MIVS} provides a principled AI-based approach to measurement validation in inverter-dominated microgrids.
In this regard, the main contributions of this paper are summarized as follows:}
\begin{itemize}
    \item \Ablack{Development of a deep-learning-assisted measurement integrity validation framework for detecting FDIAs targeting AC and DC LCDRs in islanded inverter-based microgrids, using only short windows of synchronized relay current measurements.}

    \item  To validate this framework: 
    \begin{itemize}
    \item We perform a comprehensive evaluation of the proposed \Ablack{MIVS} on multiple LCDRs, of both AC and DC types, in an inverter-based microgrid under a wide range of fault and FDIA scenarios, operating conditions, LCDR locations, sensitivity analysis, and real-time hardware-in-the-loop validation using an OPAL-RT simulator. 
    \item We conduct a comprehensive comparative analysis against existing methods, both qualitative and quantitative,  demonstrating the performance advantages of the proposed MIVS over prior art in terms of detection accuracy, applicability, power system components requirements, and detection speed.
\end{itemize}
\end{itemize}

\Ablack{The remainder of this paper is organized as follows. Section~\ref{section:Preliminaries_Threat_Model} introduces the system and threat models. Section~\ref{section:MIVS} presents the proposed measurement validation framework. Section~\ref{section:Simulation_Results} evaluates its performance through extensive case studies, followed by real-time validation in Section~\ref{section:hil}, and then comparative analysis with related work in~\ref{section:related}.} Section~\ref{section:Conclusion} concludes the paper.

\section{LCDR Operating Principle and Threat Model} \label{section:Preliminaries_Threat_Model}

\subsection{Operating Principle of LCDRs}

LCDRs are used to protect lines in inverter-based microgrids due to their 1)  sensitivity, as differential relays have the capability to detect various internal faults, including high-impedance faults, which cannot always be detected by other relays; 2)  selectivity, which is refusal to operate due to external faults and disturbances;
and 3)  speed, since LCDRs are faster than alternative schemes like directional overcurrent or distance protection
\cite{SEL_inc,DC_LCDR_ref}. These protective advantages largely arise because LCDRs operate on the fundamental principle of Kirchhoff's current law. 

\begin{figure}[t!]
\centering
\includegraphics [width=0.49\columnwidth] {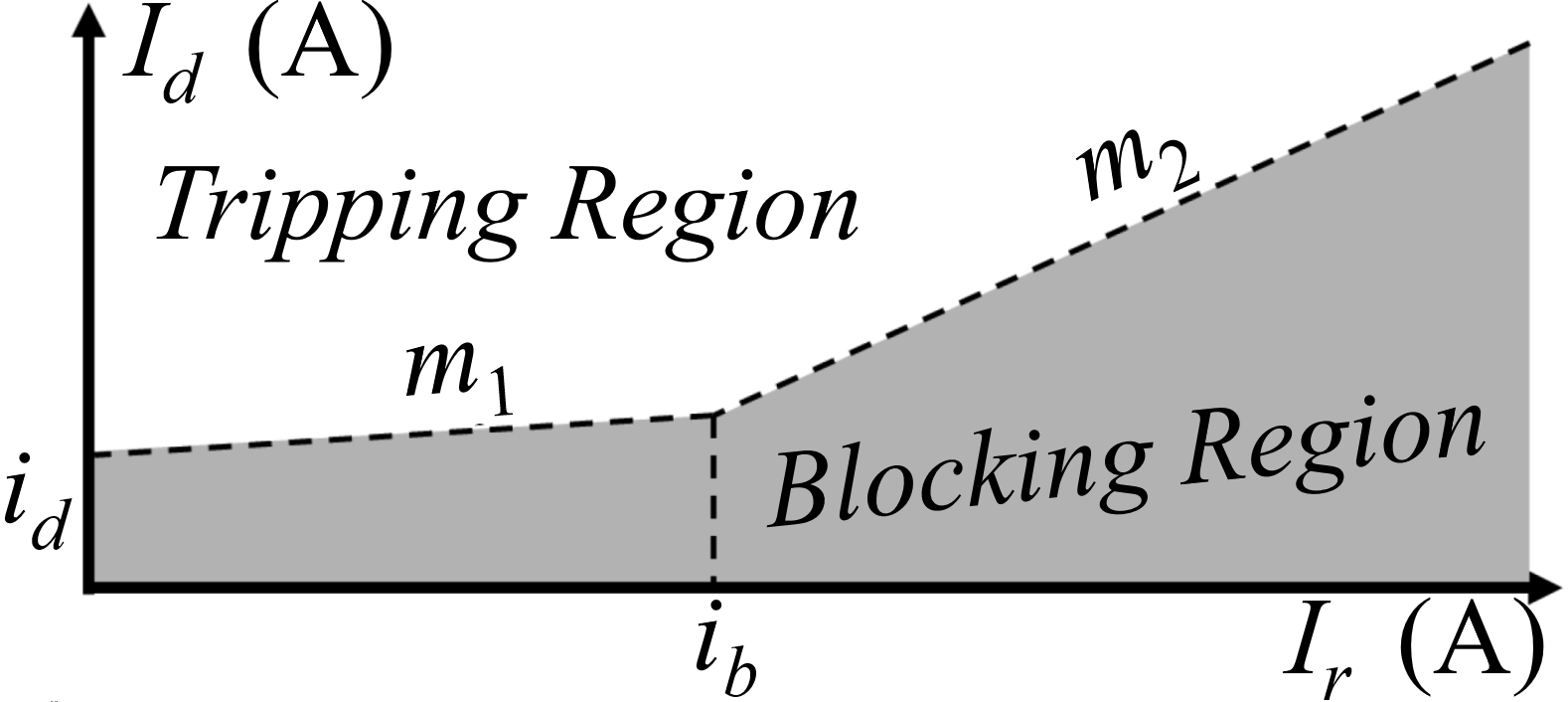}  
\includegraphics [width=0.49\columnwidth] {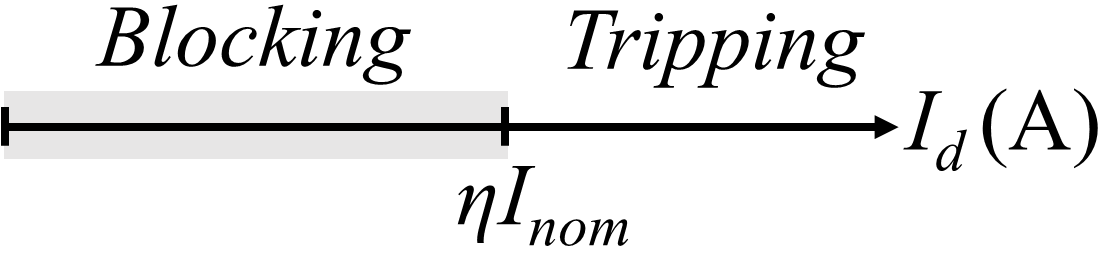}
 \\  
(a) \hspace{110pt}  (b)
\caption{LCDR's characteristics, (a) AC type, (b) DC type.}
  \label{fig:LCDR_ccs}
\end{figure}

\textit{1) AC LCDRs:}  Typically, an LCDR is located near one end of the line to be protected. To protect this line, an LCDR receives the locally-measured and remotely communicated instantaneous sampled current measurements \cite{howLCDRworks}.
From there, at each time step $k$, the LCDR's differential current ($I_d$) and restraining current ($I_r$) are calculated as

\begin{equation}
\label{eqn:diff_AC}
    I_d^{AC} [k] =|| I_1[k] + I_2[k] ||
\end{equation}

\noindent and %

\begin{equation}
    I_r [k] =|| I_1[k] || + || I_2[k] ||
\end{equation}

\noindent where $ I_1 [k]$ and $ I_2 [k]$ denote the sampled values of the local and remote currents, at time step $k$, respectively, and $||.||$ is the filtering and
magnitude estimation operation of the enclosed quantity.  
\Ablue{The current signals are first passed through standard anti-aliasing and measurement filtering stages typically employed in digital relays \cite{SEL_inc}. The magnitude estimation is performed using a sliding-window approach over the sampled current waveforms, consistent with practical relay implementations. In this work, the focus is on the resulting processed current signals, which are assumed to be time-synchronized and preconditioned as per standard LCDR measurement pipelines.}
The LCDR's operating current ($I_{op}$) is determined as

\begin{equation}
I_{op} (I_r) =                       
\begin{cases}  
   i_d + m_1 \times I_r [k]   & \text{ } I_r  [k]\leq i_b   \\
    i_d + m_1 \times i_b + m_2 \times (I_r  [k] -  i_b)    & \text{ } I_r  [k] > i_b 
\end{cases}
\end{equation}

\noindent in which $i_d$, $i_b$, $m_1$ and $m_2$ are the LCDR's operating criteria settings, as depicted in Fig. \ref{fig:LCDR_ccs} (a). The LCDR then trips   if 

\begin{equation}
\label{eqn:trip_AC}
    I_d^{AC} [k] \ge I_{op} (I_r)
\end{equation}

\noindent which indicates an internal fault on the LCDR-protected line, since under internal faults, a new path is created for the current to flow through, causing a large difference between $I_1$ and $I_2$ \cite{Saadat}.


\textit{2) DC LCDRs:}   \Ablue{In DC systems, LCDRs rely solely on local and remote current magnitudes. 
The differential current is determined as 
\begin{equation}
\label{eqn:diff_DC}
    I_d^{DC} [k] =  || I_1[k] + I_2[k] ||
\end{equation}
\noindent Herein, $I_d^{DC}$ is ideally zero under non- and external-fault conditions where  $I_1[k] \approx -I_2[k]$.} However, during faults, the following condition is satisfied

\begin{equation}
\label{eqn:trip_DC}
    I_d^{DC} [k] \ge \eta I_{nom} 
\end{equation}

\noindent in which $I_{nom}$ is the nominal current the line would carry, and $\eta$ is a pre-defined reliability threshold, as depicted in Fig. \ref{fig:LCDR_ccs} (b).  $\eta$'s value is typically between 0.1 and 0.25 \cite{DC_LCDR_ref}, and is selected to ensure that the LCDR can detect all internal faults and remains inoperative during external disturbances.

\subsection{Threat Model}

\Ablack{LCDRs represent attractive targets for cyberattacks aiming to falsely trip specific line(s) in the microgrids, disrupting the power flow and potentially inducing a system collapse. In other words, attacks on LCDRs could cause an impact similar to directly attacking microgrid switches but would remain stealthier by attacking one or more LCDRs and fooling them to trip.
We consider attackers that have access to communicated remote current measurements, in a way that allows them
to perform this false-tripping attack against a certain LCDR.} To perform this attack, a malicious entity just needs to manipulate the remote current measurements of the LCDR protecting the targeted line in a way that satisfies the LCDR's operating criteria 
\cite{Remedial_pilot_main_protection,ResMVDC}.
Manipulation of the LCDR's remote current can take different forms, some of which are simple to implement but have high impact, such as multiplying $I_2$ by a negative integer $\alpha$,
which ensures the LCDR’s operating criteria are met.
%
Alternatively, attackers can numerically solve Equations \ref{eqn:diff_AC} and \ref{eqn:trip_AC}, for AC LCDRs, or \ref{eqn:diff_DC}  and \ref{eqn:trip_DC} for DC ones, which can be found in the manufacturer's catalogs, to determine the range of $I_2$ values that would cause the targeted LCDR to trip.  
%
%
%
Adversaries can manipulate the magnitude and/or phase angle information of the LCDR's remote measurements in several ways, including by: 
\begin{enumerate}
    \item  spoofing  the GPS signal used by the attacked LCDR using noise with the same frequency  as the original GPS signal, irrespective of the LCDR's  communication media, which is equivalent to manipulating the phase angle information of the LCDR's remote measurements \cite{Remedial_pilot_main_protection}, and
    \item  intruding into the two-way communication network over which the remote measurements of the targeted LCDR are sent. This is possible because many LCDRs   rely on  communication of vulnerable media, (e.g., TCP/IP, microwave, and radio communication). Malicious entities can break into vulnerable communication media, such as microwaves and radio links, in order to eavesdrop, intercept and synthesize the communicated messages \cite{Dolev,ResMVDC}.  Additionally, the communication links have other possible network intrusion points such as communication routers and switches \cite{ResMVDC,fiber}.   
\end{enumerate}

\begin{figure}[t!]
\centering
\includegraphics
[width=1.0\columnwidth] {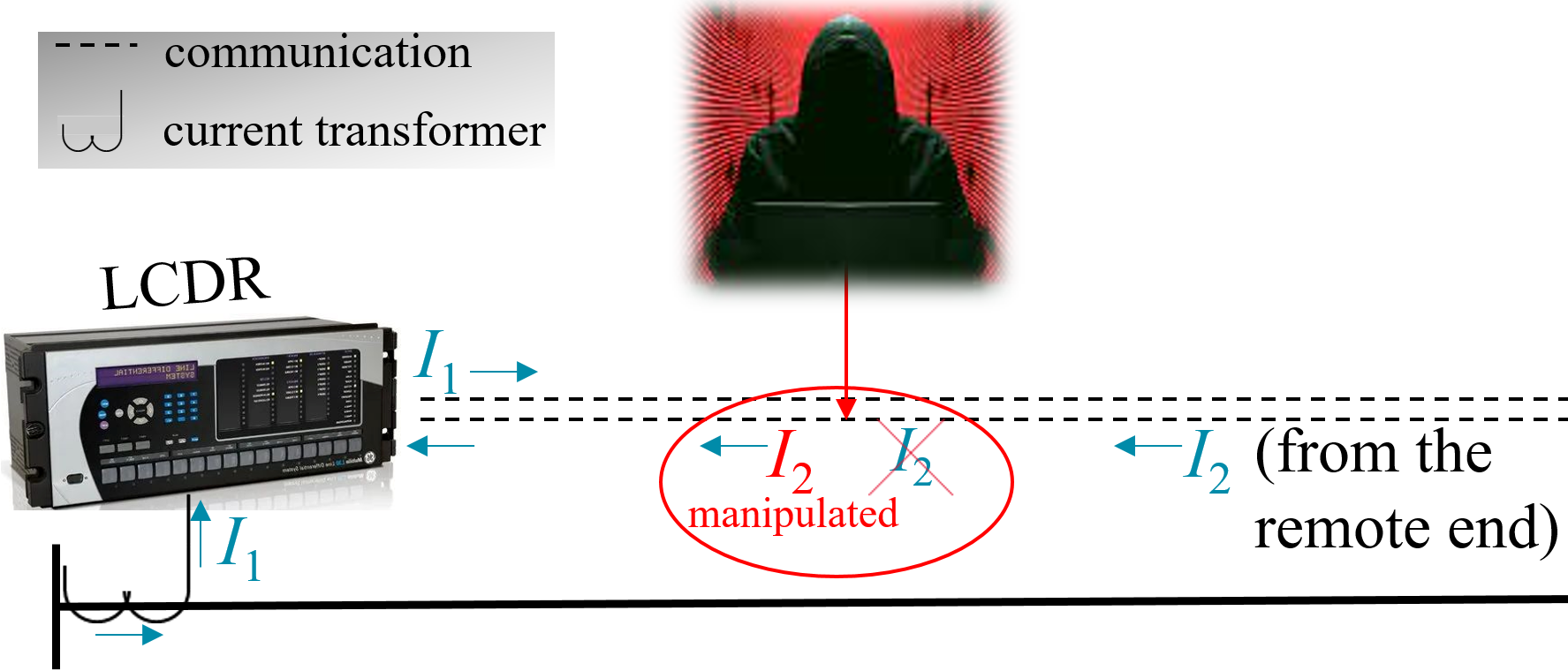}
\caption{\Ablue{Illustration of FDIA mechanisms on LCDRs. The adversary intercepts the remote current measurement $I_2$ from the communication link and manipulates its magnitude or phase to synthesize a falsified signal $I_2^m$ (manipulated $I_2$), therefore the relay receives the inconsistent pair ($I_1, I_2^m$), which satisfies tripping criteria despite the absence of a real physical fault.}}
  \label{fig:LCDR_idea}
\end{figure} 
 
Based on the above, adversaries are assumed to have basic knowledge of LCDR protection and are capable of  manipulating the magnitude and/or phase angle value of the LCDR's remote current measurements, $I_2$, as depicted in the illustration in \Ablue{Fig. \ref{fig:LCDR_idea}, which illustrates an FDIA attack scenario targeting an LCDR. In other words, the relay receives the locally-measured current $I_1$ directly from the current transformer at the relay's terminal, via a short copper wire not susceptible to remote manipulation. The remote current $I_2$, measured at the far end of the protected line, is communicated to the relay over a digital network. An adversary with access to this communication channel can intercept and replace the legitimate $I_2$ with a manipulated value $I_2^m$, crafted to satisfy the LCDR's tripping criteria (Equations~1 and~4 for AC LCDRs, or Equations~5 and~6 for DC LCDRs) in the absence of any real internal fault. The relay, unable to distinguish between a genuine fault-induced differential current and a cyber-manipulated one using its classical operating logic alone, issues a false trip command, which is precisely the scenario that the proposed MIVS is designed to prevent.}

On the other hand, in this threat model, local measurements are generally considered secure from manipulation as they can be directly sent from the current transformers to the LCDR via copper wires where there is no room to manipulate them.
Based on the above, FDIAs on AC LCDRs can be modeled as

\begin{equation}
    I_d^{AC, m}   [k] =|| I_1[k] +   I_2^m [k]    ||
\end{equation}

\noindent where $I_2^m$ is the manipulated remote $I$ measurement. 
 Given that the value of $ I_d  [k]$ is normally zero, rewriting (1) yields

\begin{equation}
             I_1[k] = - I_2[k] 
\end{equation}
\noindent which if substituted into (7) yields 

\begin{equation}
    I_d^{AC, m}   [k] =|| -I_2[k] +   I_2^m [k]    ||
\end{equation}

\noindent for AC LCDRs, and for DC LCDRs 

\begin{equation}
    I_d^{DC, m}   [k] =|| -I_2[k] +   I_2^m [k]    ||
\end{equation}
%

%
\noindent For instance, adversaries can obtain the value of  $I_2$ by eavesdropping, then manipulate the remote current measurement of any of the three phases so that $I_2^m [k]$ satisfies the LCDR's operating criteria explained above. 
%
Further, adversaries can perform  time-synchronization attacks (TSAs) against the LCDR, e.g., manipulate the time-synchronization mechanism used by the targeted LCDR (e.g., GPS signal) which is equivalent to manipulating  the phase angle of the LCDR’s remote current, resulting in false tripping of the targeted LCDR \cite{ResMVDC,Remedial_pilot_main_protection}. TSAs are also studied in this paper. 

It is worth noting that this paper focuses on cyberattacks that can be launched remotely. Consequently, denial-of-service attacks
\cite{ResMVDC}, including those that aim to prevent an LCDR from operating during actual faults or disrupt its communication link, are not explicitly considered. This exclusion is justified by two key factors: 1) modern LCDRs are typically equipped with redundant communication links and digital mechanisms for detecting and responding to communication failures \cite{SEL_inc}; and 2) it is very difficult for remote attackers to precisely time their cyberattack with the occurrence of an actual fault, particularly because remote attackers cannot predict the exact inception time of faults and typically aim to minimize the duration of their activity to avoid detection.
%

\section{Developing an \Ablack{MIVS} for LCDRs in Islanded Inverter-Based Microgrids} \label{section:MIVS}

\subsection{Solution Requirements}

The problem posed by FDIAs on LCDRs  is the false tripping of the LCDR-protected line under no real fault. Attacking more than one LCDR (and hence taking down more than one line) after one another can cause microgrid instability or shutting down. 
\Ablack{This problem can be mitigated if the FDIA is appropriately detected by validating the LCDR measurements once LCDR is triggered.} 

\begin{figure}[t!]
\centering
\includegraphics
[width=1\columnwidth] {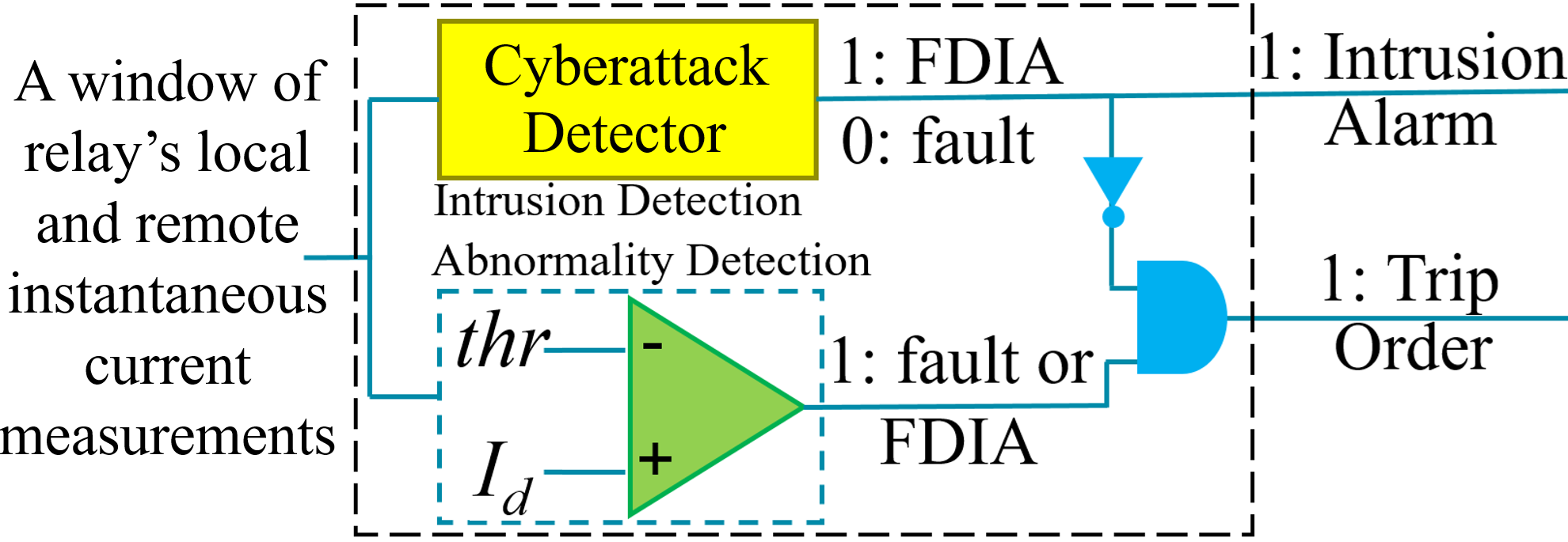}
\caption{ \Ablack{LCDRs augmented by the proposed \Ablack{MIVS}.  $thr$ denotes threshold, i.e., $I_{op}$ and $\eta I_{nom}$ for AC and DC LCDRs, respectively.}}
  \label{fig:scheme}
\end{figure}

\Ablack{To tackle this problem, Fig. \ref{fig:scheme} illustrates the information flow of LCDRs secured using the proposed \Ablack{MIVS}.
The proposed scheme operates as a supervisory validation layer and does not modify the underlying differential protection logic or thresholds.}
The proposed \Ablack{MIVS} is trained offline on various faults and FDIAs. When online, once an LCDR is triggered to trip (i.e., by a fault or an FDIA), the \Ablack{MIVS} must be able to confirm the occurrence of a fault before allowing the LCDR to trip. 
 Additionally, for practicality, the \Ablack{MIVS} should rely only on the measurements available for LCDRs.  This \Ablack{MIVS} must also not significantly increase the total time that LCDR will take to detect and confirm actual faults, i.e., to maintain the LCDR's speed merit and remain within LCDR's maximum operating time limits.

A unique attribute of the proposed \Ablack{MIVS} compared to previous work is that it operates directly on the instantaneous relay current measurements recorded for window of a few milliseconds, pre and post LCDR triggering. This window of current measurements as treated as a multi-dimensional time series. The \Ablack{MIVS} can capture subtle patterns that characterize true faults and cannot be easily replicated by attackers leveraging the interphase and time dependencies of LCDR current measurements during faults vs FDIAs.
This approach can be applied to both AC and DC LCDRs in inverter-based microgrids.
The proposed approach is also different from previous works that mainly compare the LCDR current magnitudes before and after triggering to detect FDIAs  e.g., in \cite{Ahmad_smartgrid}. This is because, in inverter-based microgrids,   current magnitudes, do not necessarily change during all fault conditions, and a fortiori under FDIAs where only remote measurements are manipulated, as discussed earlier.  
Another feature that previous schemes relied on is the current phase angle measurements, while DC LCDR currents do not have phase angles.  
Further, another advantage of the proposed approach is eliminating the classical phase of extracting hand-crafted feature values from the instantaneous current measurements (required in most previous techniques), thus reducing the time complexity of the proposed \Ablack{MIVS}.

 \subsection{Recurrent Neural Networks for FDIAs Detection in LCDRs }
This paper leverages RNNs as a tool in the proposed \Ablack{MIVS}. This tool can be trained offline on only relay measurements under both faults and FDIAs, and used online to detect possible FDIAs that manipulate relay measurements aiming to falsely trip the line protected by the targeted LCDR.
RNNs are known for their ability to learn unique patterns of hierarchical and discriminative features directly from raw time series data, i.e., instantaneous relay current measurements \cite{Williams_RNN}. These merits make RNNs suitable for operating on the current measurements of the LCDR represented as time series data.
%
%
 RNNs differ from regular neural networks in that RNNs do not assume that the input data are independent.  They predict an output by learning the temporal correlation between inputs \cite{Williams_RNN}, which is more suitable for our problem compared to other models, such as CNN, which treats time as spatial, and LSTM adds complexity without measurable gain \cite{tsai, Williams_RNN}. 
RNNs develop a memory construct to compute the new output based on previous output information. 

\begin{figure}[t!]
\centering
\includegraphics
[width=1.0\columnwidth] {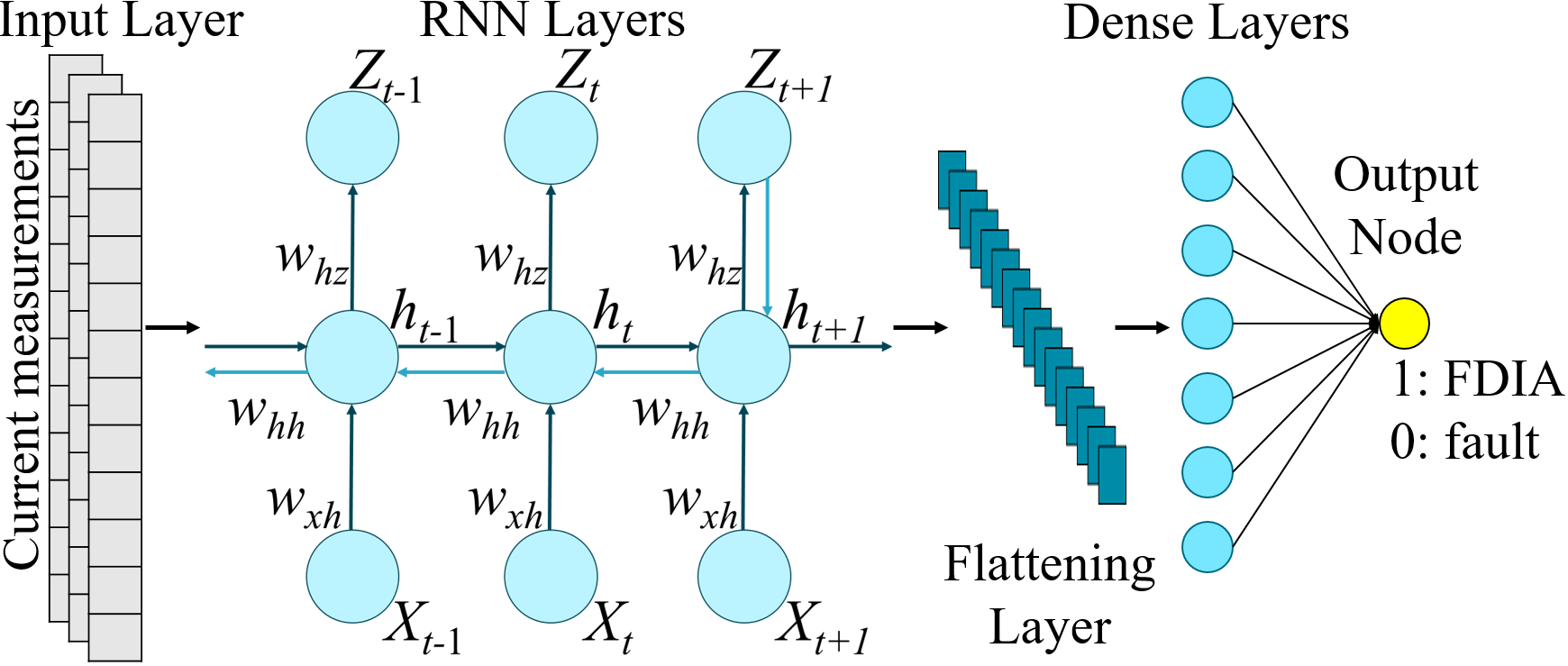}
\caption{ Architecture of the proposed RNN-based \Ablack{MIVS}.}
  \label{fig:RNN_Arch_Complete}
\end{figure}

\Ablue{For an observation sequence $X_i, X_{i+1}, \ldots, X_T \in \mathbb{R}^d$,
where $d$ is the number of input measurement channels (i.e., $d = 6$ for AC LCDRs with
three-phase local and remote currents, and $d = 4$ for DC LCDRs with two-pole local and
remote currents), and whose corresponding labels are $y_i, y_{i+1}, \ldots, y_T \in
\{0, 1\}$ (with 0 denoting a genuine internal fault and 1 denoting an FDIA), the goal
of training an RNN model is to find a non-linear mapping function $f$ that maps the
input sequence to its corresponding label, i.e., fault or FDIA} \cite{tsai}.  
RNNs send feedback signals to process time-dependent data, making subsequent outputs dependent on computed output. This is also known as the hidden state and can be represented as

\begin{equation}
    h_t = f  (h_{t-1} , X_t)
\label{equation:h_t}
\end{equation}

\noindent in which $h_{t-1}$ is the hidden state at time $t-1$ and
$X_t$ represents the multi-feature input at time step $t$. 
Often, equation (\ref{equation:h_t}), which is the essence of RNNs, is computed as:

\begin{equation}
    h_t = \text{tanh} \left(W_{hh}\text{ }h_{t-1} + W_{xh}\text{ }x_t\right)
\end{equation}

\noindent where $W_{hh}$ and $W_{xh}$ are the weight matrices. The hyperbolic tangent function, tanh, introduces non-linearity into the hidden state computation process. The RNN output is denoted $z_t$. The final hidden state $h_T$ serves as a representation of the sequence \cite{RNN_tutorial}.  The architecture of RNNs is shown in Fig. \ref{fig:RNN_Arch_Complete}.  

\noindent Afterward, $h_T$ is  passed through a softmax-activated dense classification layer, which yields a predicted class probability, $\hat{Y}$, as follows:

\begin{equation}
    \hat{Y}_i = \frac{e^{(W_{hy}\text{ }h_T)_i}}{\sum_{j=1}^C e^{(W_{hy}\text{ }h_T)_j}}
\end{equation}

\noindent where $W_{hy}$ is the output layer's weight matrix, and $C$ is the  number of classes. Given the binary nature of our problem, \Ablue{the RNN is trained using binary cross-entropy loss function:
\begin{equation}
    \mathcal{L} = - \frac{1}{N} \sum_{k=1}^{N} \left[ y_k \log(\hat{y}_k) + (1 - y_k)\log(1 - \hat{y}_k) \right]
\end{equation}
where $N$ denotes the number of training samples in the batch,  $y_k \in \{0,1\}$ is the true label of the $k$-th sample, and $\hat{y}_k \in (0,1)$ is the RNN-predicted probability that the sample is an FDIA.}
With the ability to learn from multiple input channels, an RNN model can be trained both on the local and remote current measurements of the LCDRs, one channel per phase/pole current measurement. 

%
\section{Performance Evaluation of the proposed \Ablack{MIVS}} \label{section:Simulation_Results}
  This section presents the performance evaluation procedure of the \Ablack{MIVS} and presents the results of different case studies.  
  %
%
%
\subsection{Test Microgrid} \label{section:Test_System}
\begin{figure}[t!]
\centering
\includegraphics
[width=1.0\columnwidth] {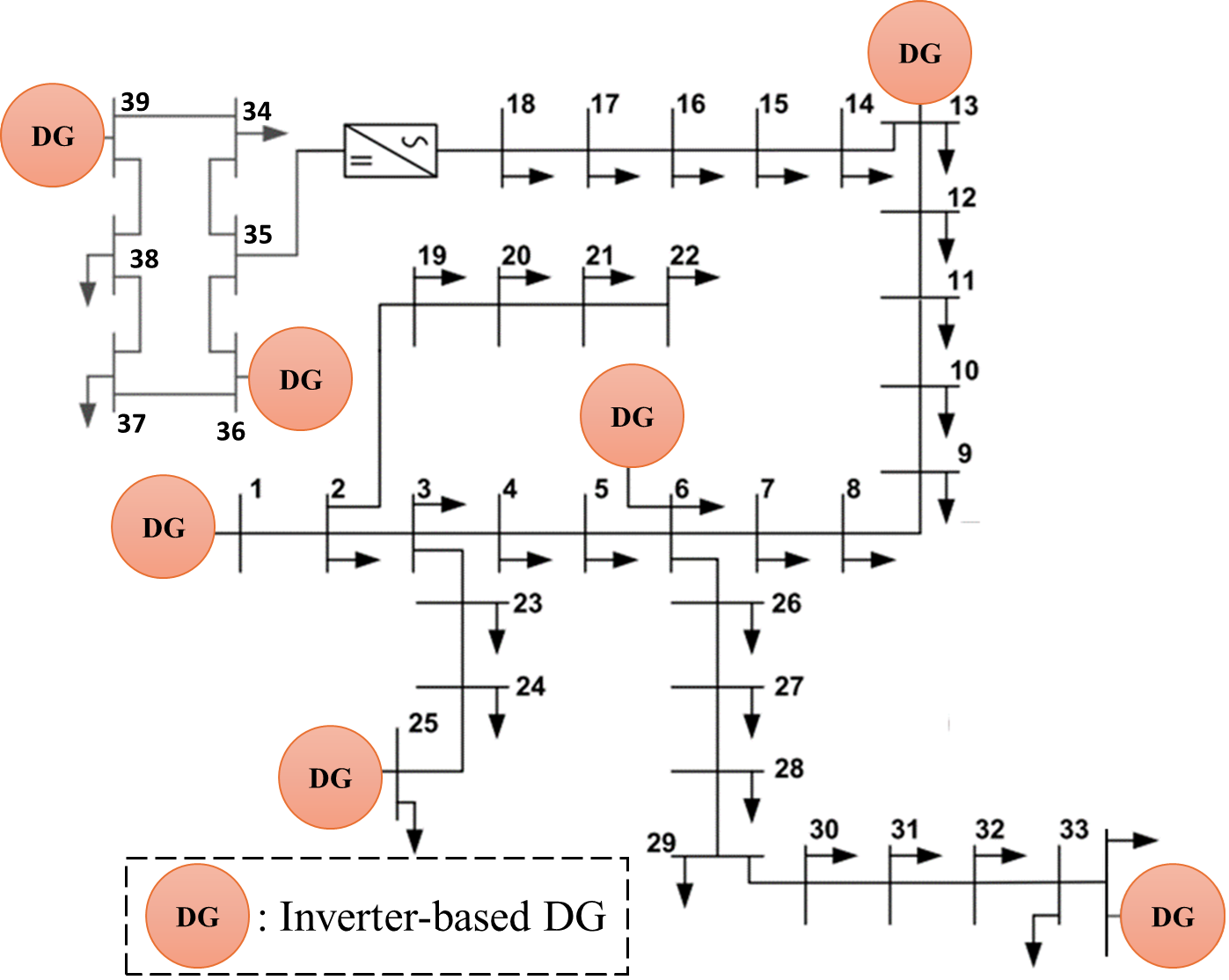}
\caption{ Test system.}
  \label{fig:test_system}
\end{figure}

The performance of the proposed \Ablack{MIVS} approach is evaluated using the medium-voltage inverter-based microgrid test system depicted in Fig.   \ref{fig:test_system}, 
simulated in PSCAD/EMTDC environment.
This \Ablack{inverter-based microgrid} has an AC side which is based on the IEEE 33-bus distribution benchmark system and is interconnected, via a bidirectional interlinking AC-DC converter, with a 6-bus DC side, as illustrated in the Figure. 
 In the test system, all DGs are droop-controlled and are interfaced with the microgrid via DC-to-AC inverters and DC-to-DC converters on the AC and DC sides, respectively \cite{Testsystem_Holomorphicpaper}.
%
%
%
%
%
%
Additionally, all DGs are equipped with hard fault current limiters, which are self-protection mechanisms employed by modern DGs, that cap the current withdrawn from each DG during fault conditions. These limiters are set to allow a maximum of 1.5 p.u of the DG's current to flow into the inverter-based microgrid during faults. 
The test system's total active and reactive power demands are 12.35 p.u. and 4.6 p.u., respectively, for a base complex power of 1 MVA.
In the test inverter-based microgrid, any line can be potentially protected by LCDRs for accuracy and speed needs.  LCDRs settings $i_d$, $i_b$, $m_1$, $m_2$, and $\eta$  are set as  0.05 kA, 0.585 kA, 0.2, 0.4, and 0.2, respectively, following
\cite{ SEL_inc,DC_LCDR_ref}. 

\subsection{Performance Evaluation Metrics}

The performance of the proposed approach is evaluated in several case studies using the following standard metric: 

\begin{equation}
   \text{Accuracy} = \frac{\text{TPs + TNs}}{\text{TPs + TNs + FPs + FNs}} 
\end{equation}
\vspace{2pt}

\noindent where True Positives (TPs) and   False Negatives (FNs) are detected and undetected FDIAs, respectively, while True Negatives (TPs) and False Positives (FPs) are correctly classified and misclassified faults, respectively.  
For thorougher analysis, Precision, Recall, and F1-score metrics are used, defined as

\begin{equation}
    \text{F1-score} = \frac{2 \cdot \text{Precision} \cdot \text{Recall}}{\text{Precision} + \text{Recall}},
 \quad \text{where} \end{equation}

 \begin{equation*}
\text{Precision}  = \frac{\text{TPs}}{\text{TPs} + \text{FPs}} 
\textcolor{white}{..} (20) \textcolor{white}{......}
   \text{Recall} = \frac{\text{TPs}}{\text{TPs} + \text{FNs}}
\textcolor{white}{..} (21)
\end{equation*}
\subsection{Results of Comprehensive Case Studies 1 and 2} \label{section:LCDR_1_LCDR_2}

\Ablack{In this subsection, we showcase the performance of the proposed \Ablack{MIVS}  by focusing on two representative LCDRs as case studies. Specifically, we train and evaluate one \Ablack{MIVS} for an AC LCDR, $MIVS_{AC}^{1-2}$, and another for a DC LCDR, $MIVS_{DC}^{34-35}$. The selected LCDRs are $LCDR_{1-2}^{AC}$, which protects the AC line 1-2 near bus 1, and $LCDR_{34-35}^{DC}$, which protects the DC line 34-35 near bus 34. Each \Ablack{MIVS} is trained to differentiate between faults and FDIAs affecting its corresponding LCDR.}

\textit{1) FDIA and Fault Evaluation Scenarios:}  
\Ablack{To ensure comprehensive evaluation, we simulate a wide range of fault and FDIA scenarios for each LCDR. }
On one hand, a wide spectrum of fault cases are simulated considering different fault parameters, as follows:

\begin{itemize}
    \item \textit{ Fault type:}   For $LCDR_{1-2}^{AC}$, which protects a line with three phases denoted $A$, $B$ and $C$,  the following fault types are simulated: $A$-$G$, $B$-$G$, $C$-$G$, $A$-$B$, $B$-$C$, $C$-$A$, $A$-$B$-$G$, $B$-$C$-$G$, $C$-$A$-$G$,  $A$-$B$-$C$, and  $A$-$B$-$C$-$G$, where $G$ denotes the ground. Similarly, for $LCDR_{34-35}^{DC}$ protecting a DC line with two identical cables, one for the positive pole ($PP$) and one for the negative pole ($NP$), the following fault types are simulated:  $PP$-$NP$, $PP$-$G$, and $NP$-$G$.

    \item \textit{Fault impedance:}  Different fault impedance values ($Z_f$) are simulated, where $Z_f$ $\in$ [0,200] $\Omega$, to cover  scenarios ranging from bolted to high impedance faults. 
    
    \item \textit{Fault location:}  Different fault locations, ranging from faults located at 10\% of the line's length (calculated from the LCDR's location) up to 90\% of the line length, with a step of 10\%, are simulated.

    \item \textit{Fault inception time:}  To take into consideration the effect of the fault inception angle $ \theta $,  fault starting times are varied with $\theta$ $\in$ [$0^{\circ}$, $180^{\circ} $] for $LCDR_{1-2}^{AC}$.
\end{itemize}

\noindent \Ablack{In total, 1000 fault scenarios are simulated for $LCDR_{1-2}^{AC}$ and 600 fault scenarios for $LCDR_{34-35}^{DC}$. On the other hand, 1000 FDIAs and TSAs are similarly generated for the AC LCDR and 600 for the DC LCDR, to establish a balance with fault scenarios. The FDIAs and TSAs are crafted by manipulating the remote current measurement ($I_2$) in each case to satisfy the tripping conditions of the respective LCDRs, using equations 
(1)-(10) outlined in Section \ref{section:Preliminaries_Threat_Model}.
}
These attacks are implemented through different ways of manipulating the magnitude and phase angle information of the targeted LCDR's remote measurements, including by adding to or multiplying the nominal value of $I_2$, ensuring diverse attack scenarios. 
Furthermore, to account for microgrid operational dynamics, the simulations also include cases where FDIAs occur simultaneously with dynamic system events like the connection or disconnection of DGs or loads, under different system loading conditions. This comprehensive approach ensures that the trained \Ablack{MIVS} models can effectively distinguish between legitimate faults and cyberattacks under various microgrid conditions. In each fault or FDIA scenario, the LCDR's local and remote measurements are recorded for a few milliseconds. \Ablue{A fixed-length observation window of $T$ time steps is used as input to the RNN, where $T$ is selected to capture the transient characteristics of current waveforms while maintaining low detection latency. In this work, the input observation window for $\text{MIVS}_{AC}^{1-2}$ spans 10\,ms, centered around the LCDR triggering event. For $\text{MIVS}_{DC}^{34-35}$, a shorter window of 4\,ms is used, reflecting the faster transient dynamics of DC fault waveforms. In both cases, the window is centered on the triggering instant, incorporating measurements recorded immediately before and after the LCDR pickup.}
\Ablack{The datasets for $LCDR_{1-2}^{AC}$ and $LCDR_{34-35}^{DC}$ are labeled and split into training and testing sets according to the following procedure. The fault scenarios for each LCDR are shuffled and randomly split into 80\% for training, and 20\% for testing the performance of the respective \Ablack{MIVS}.
}

\begin{figure}[t!]
\centering
\includegraphics
[width=0.49\columnwidth] {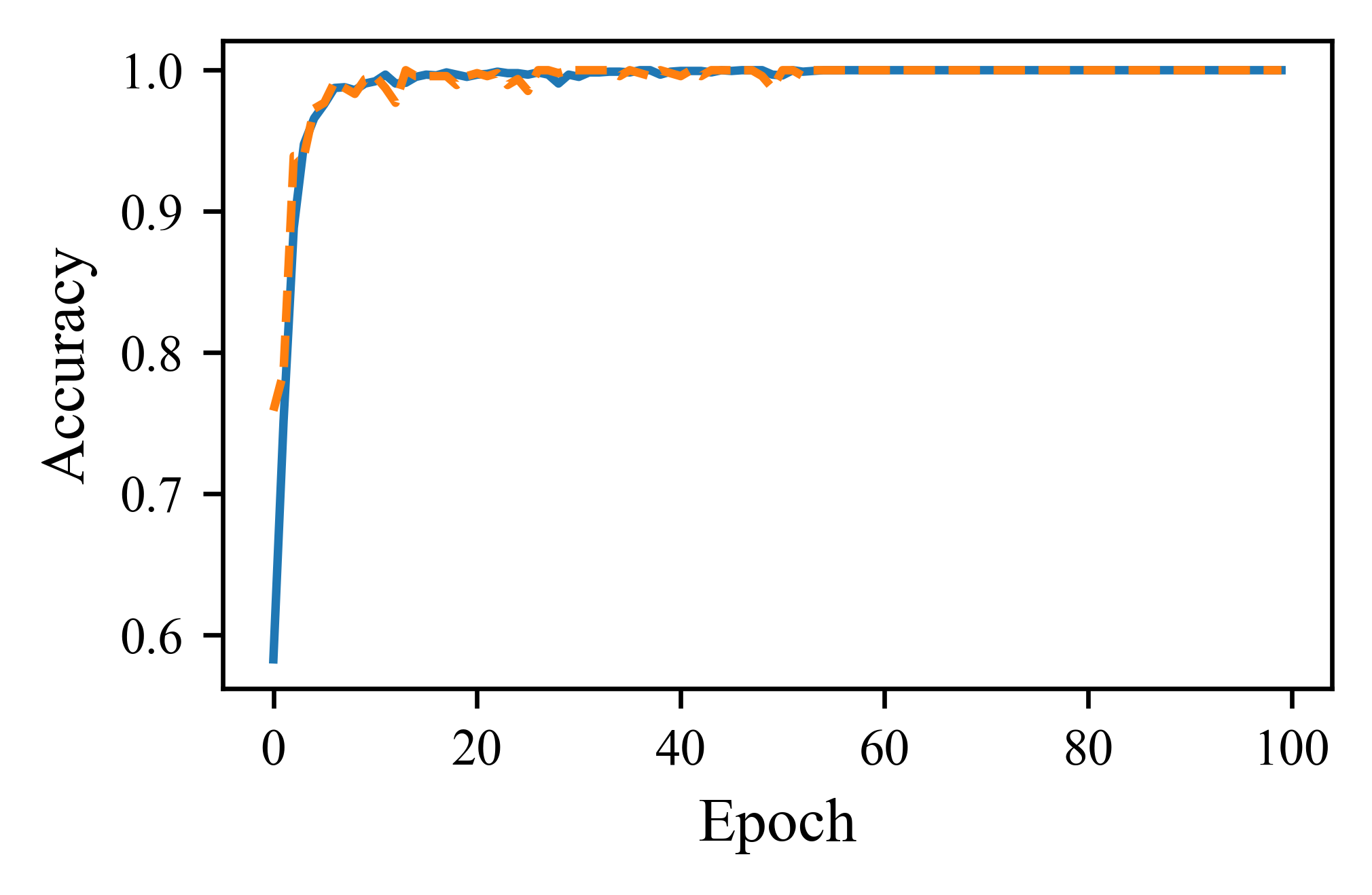}
\includegraphics
[width=0.49\columnwidth] {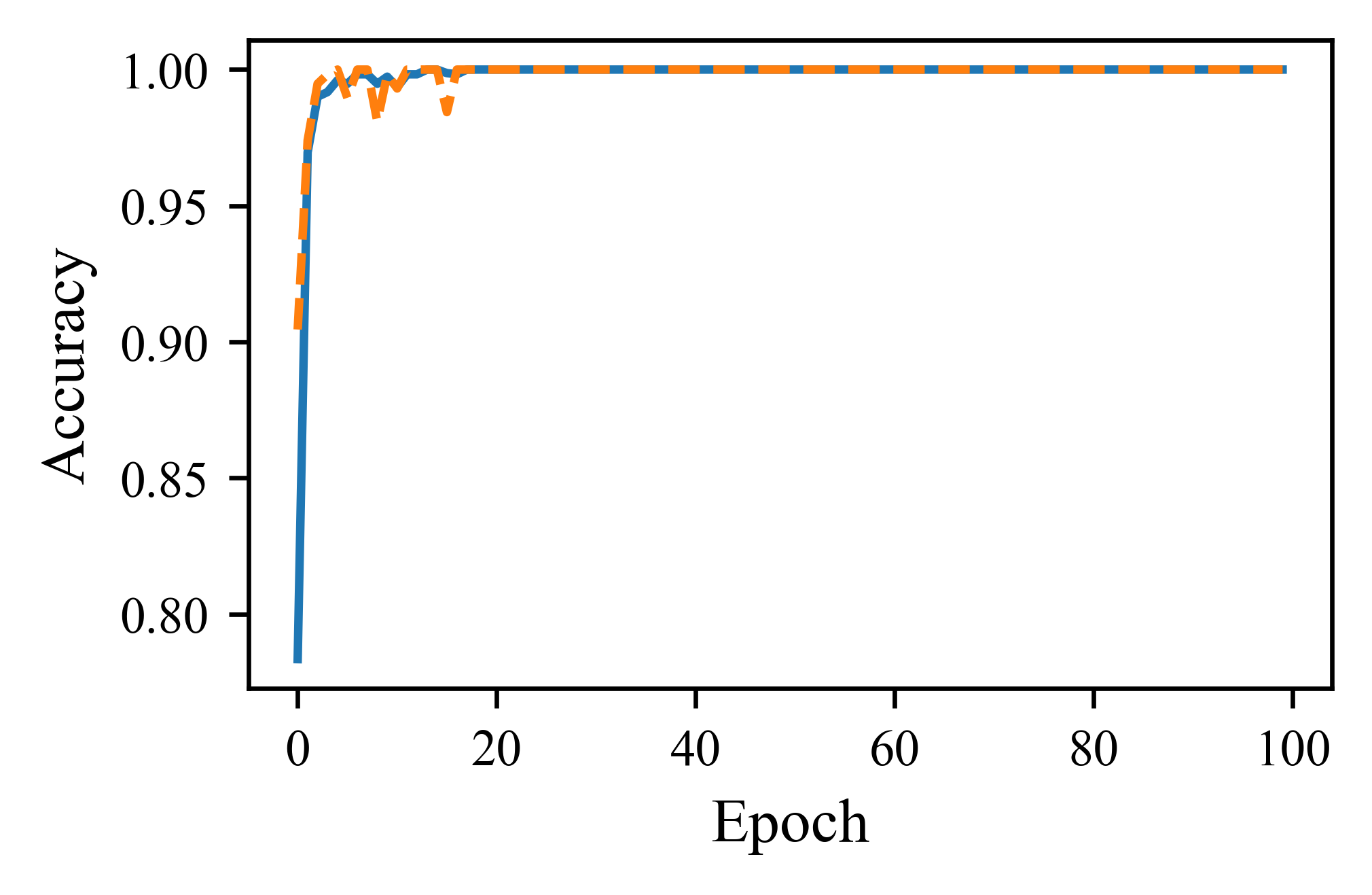}
\caption{\Ablue{Training and validation Accuracy curves over epochs for  (a)  $MIVS_{DC}^{34-35}$ and (b)  $MIVS_{AC}^{1-2}$, blue denotes training while orange denotes validation.}}
  \label{fig:training_curves}
\end{figure}

\textit{2) \Ablack{MIVS} Model Settings and Training:} 
\Ablack{In this paper, each \Ablack{MIVS}, including $MIVS_{AC}^{1-2}$ and $MIVS_{DC}^{34-35}$, employs an RNN model, as explained in Section \ref{section:MIVS}, with 3 recurrent neural network layers, followed by flattening layer, two dense layers, and a classification layer. This  model has 14,817  trainable parameters and occupies a memory space of 57.88 KB. 
The details of the above model, including number of layers are determined in a systematic way using the standard technique of automatic hyperparameter tuning, during which the model’s variables are automatically updated during the training process, with the objective of maximizing the model's accuracy, using grid search as an optimization technique \cite{Bayes}. 
Within each optimization iteration, the RNN is trained over several epochs using Adam optimizer and the Back Propagation Through Time algorithm \cite{RNN_tutorial}.
} 
\Ablue{To further illustrate the convergence behavior and learning stability of the proposed models, training and validation accuracy curves over epochs are included for both $MIVS_{AC}^{1-2}$ and $MIVS_{DC}^{34-35}$, respectively, as shown in 
Fig.~\ref{fig:training_curves}.
Both models converge smoothly within a few epochs, with training and validation curves tracking closely throughout, indicating stable learning and that the selected model architecture and training configuration are well-suited for this classification task.}

\begin{figure}[t!]
\centering
\includegraphics [width=0.49\columnwidth] {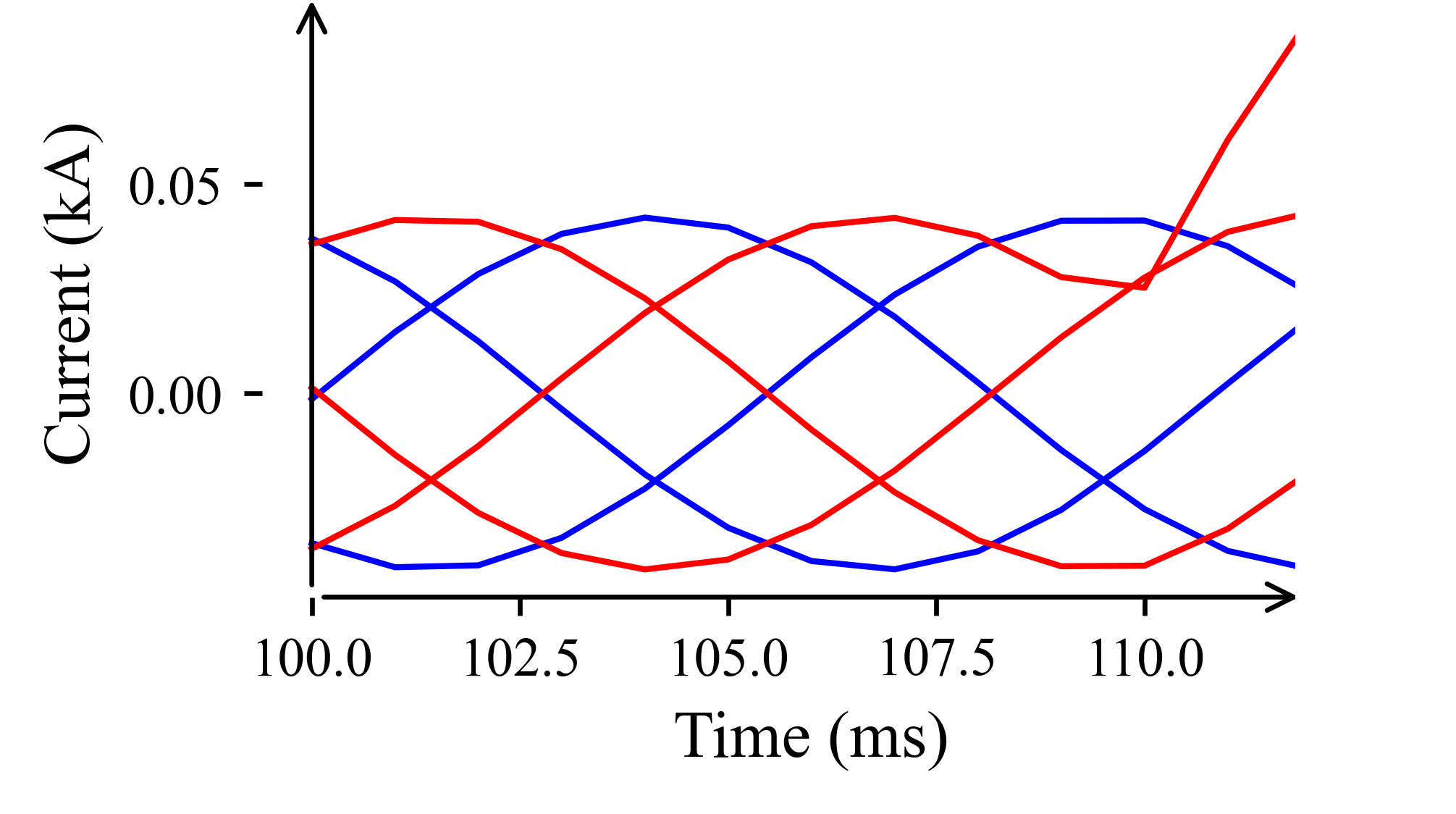}
\includegraphics [width=0.49\columnwidth] {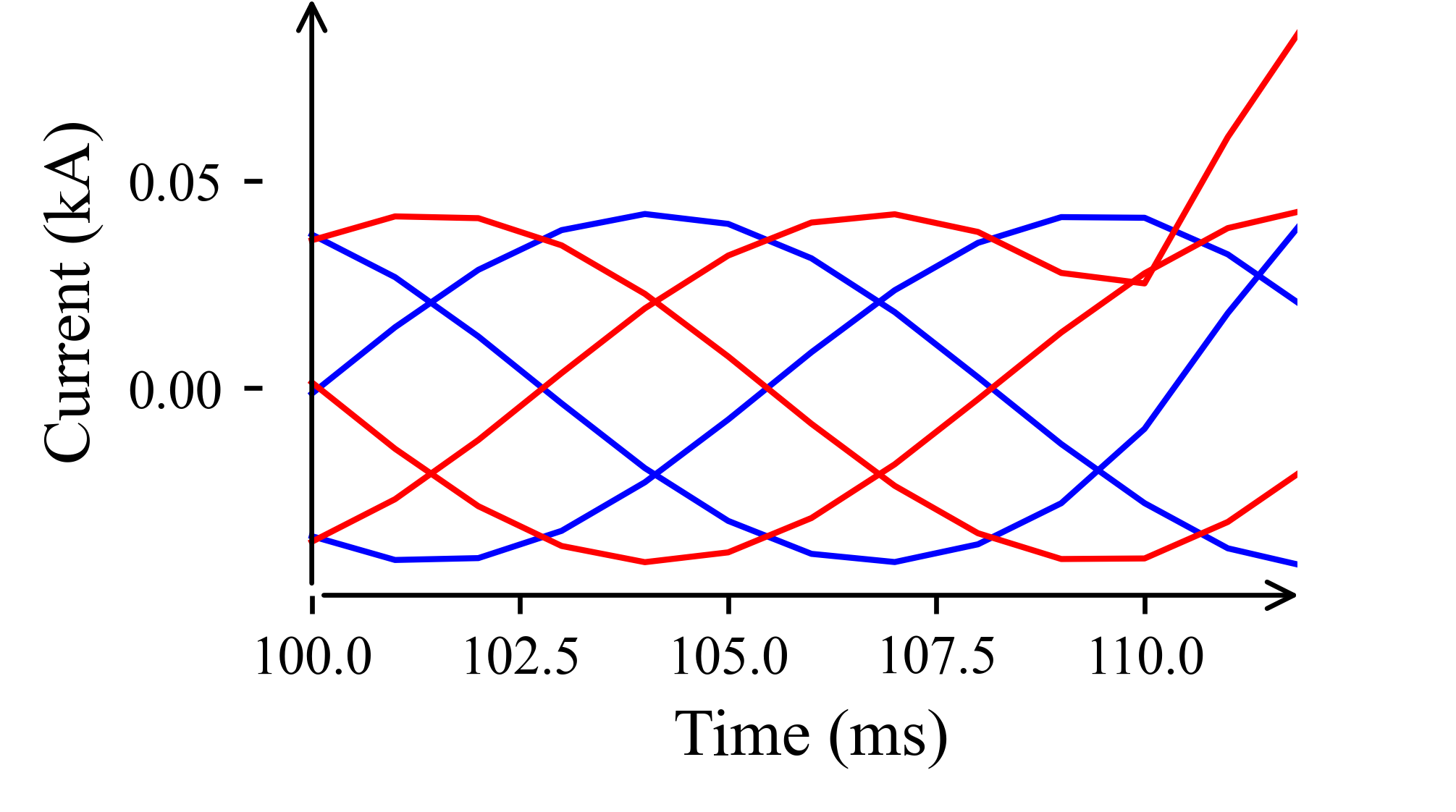}
\\ (a) \hspace{100pt} (b)
\\
\includegraphics [width=0.49\columnwidth] 
{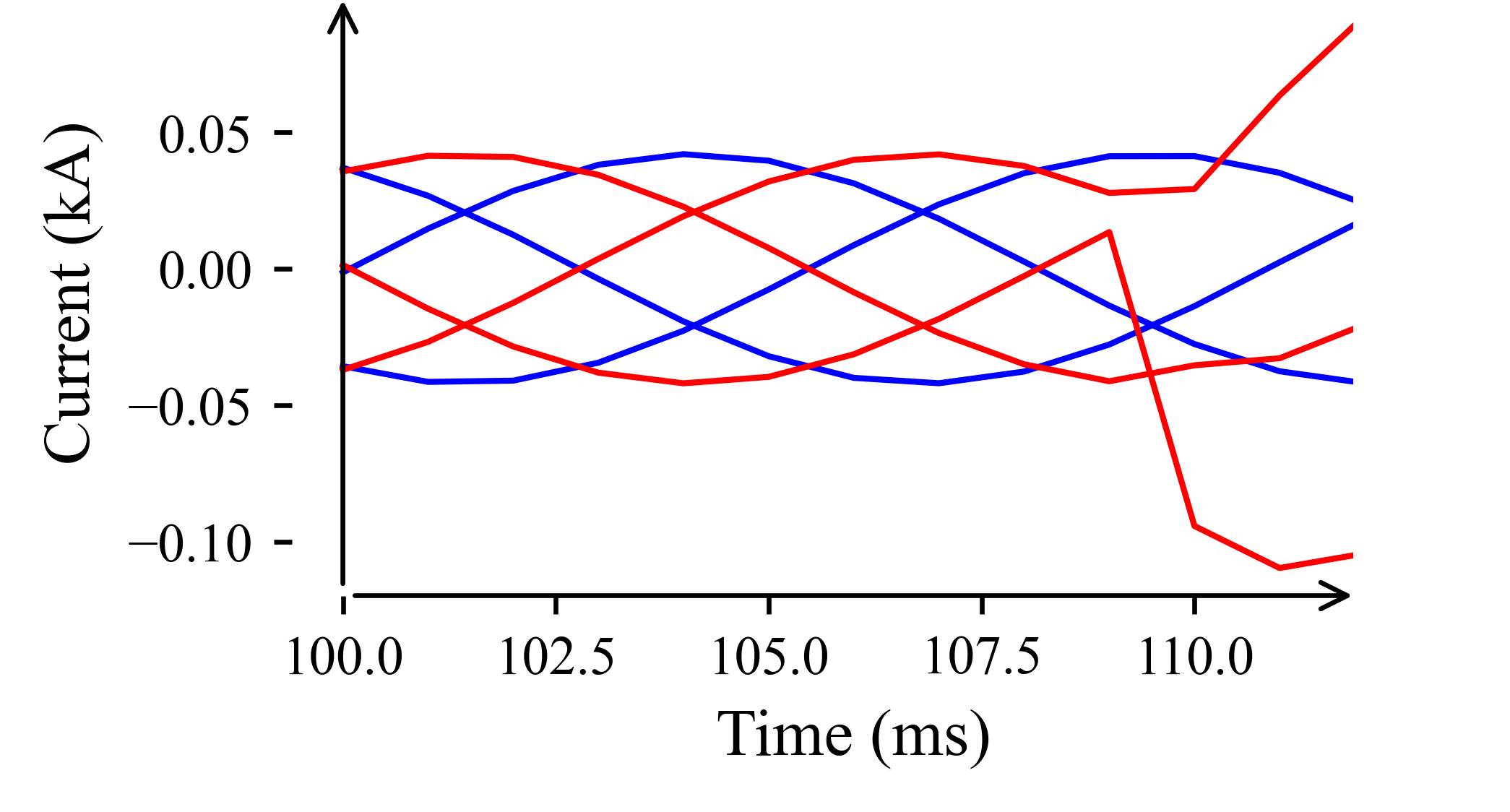}
\includegraphics [width=0.49\columnwidth] {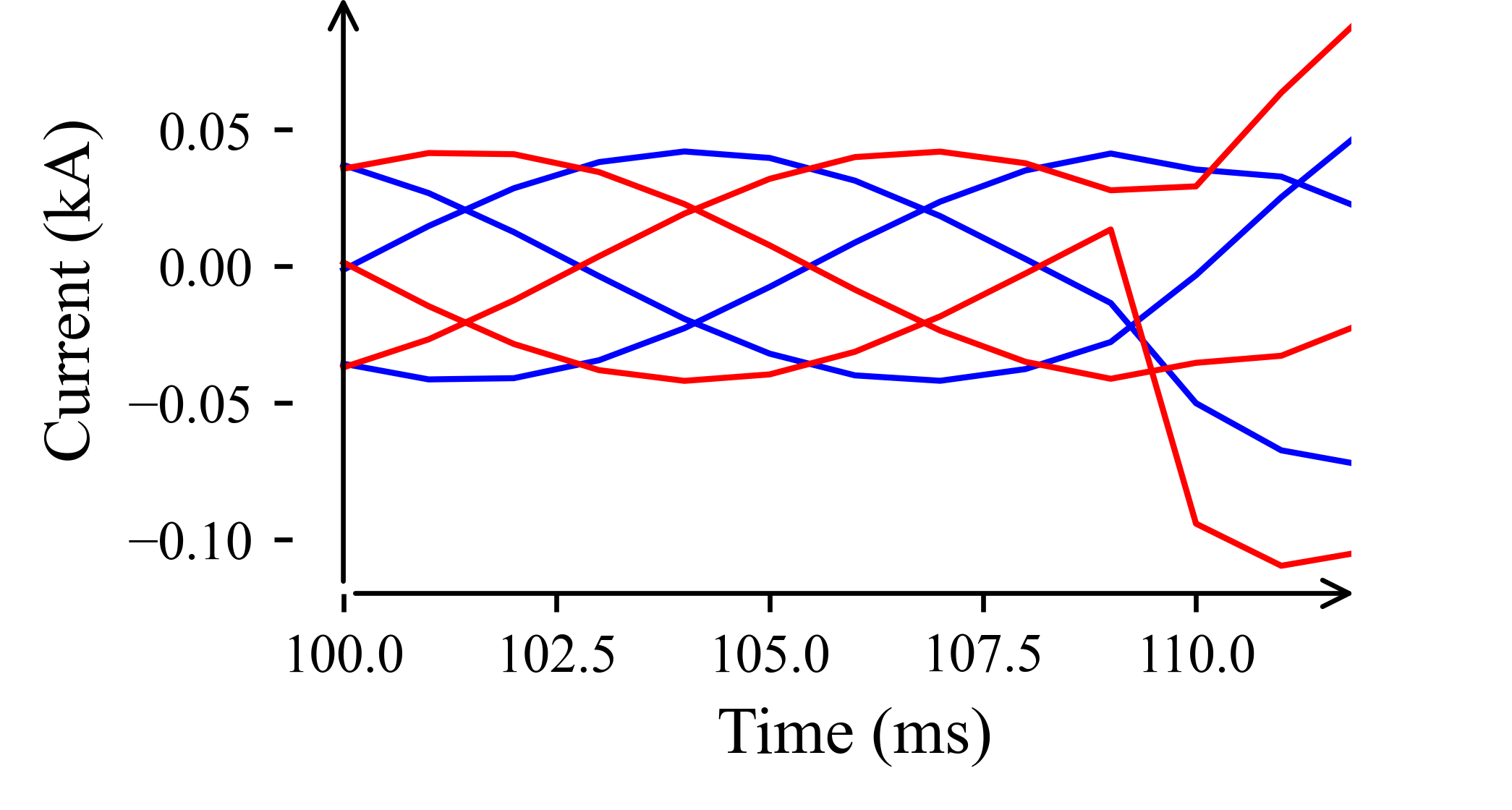}
\\ (c) \hspace{100pt} (d)
\caption{ \Ablue{Examples of FDIAs and faults investigated in this paper. Red denotes remote measurements and blue denotes local measurements. (a) an FDIA manipulating one of the 3 phase remote measurements, (b) a single-phase-to-ground fault, (c) an FDIA manipulating multiple remote current measurements, (d) measurements under a fault involving multiple phases.}}
  \label{fig:samples}
\end{figure}

\textit{3) Testing Results of $MIVS_{AC}^{1-2}$ and $MIVS_{DC}^{34-35}$:} 
Tables \ref{tab:Confusion_AC_DCR} and \ref{tab:Performance_AC_DC} summarize the testing results of the proposed cyber-resilient validation scheme for the AC and DC line current differential relays, $MIVS_{AC}^{1-2}$ and $MIVS_{DC}^{34-35}$.
From a protection perspective, the primary objective of the proposed scheme is to prevent false tripping caused by false data injection attacks while preserving relay dependability during genuine internal faults. As shown in Table \ref{tab:Confusion_AC_DCR}, $MIVS_{AC}^{1-2}$ successfully blocks 98.3\% of FDIA cases, thereby preventing erroneous LCDR operations, while only 1.7\% of FDIA instances are misclassified as faults. More importantly, 99.5\% of internal faults are correctly validated, indicating that the proposed scheme introduces negligible risk of fault non-tripping and thus preserves the inherent dependability of $LCDR_{1-2}^{AC}$.
Similarly, $MIVS_{DC}^{34-35}$ demonstrates strong protection performance by correctly identifying 99.17\% of FDIA cases and achieving 100\% correct validation of internal faults, ensuring that no legitimate fault is blocked. This result confirms that the proposed validation layer does not compromise fault clearance in DC line protection, which is critical in inverter-dominated microgrids.
In addition to these protection-oriented outcomes, the overall classification performance is summarized in Table \ref{tab:Performance_AC_DC}, where $MIVS_{AC}^{1-2}$ and $MIVS_{DC}^{34-35}$ achieve accuracies of 98.9\% and 99.6\%, respectively. These metrics further confirm the ability of the proposed \Ablack{MIVS} to reliably distinguish between faults and cyber-induced false measurements.
\Ablue{It can be observed from the results that some FDIAs are misclassified. 
For instance,  Table~I shows that a small number of FDIAs are misclassified as faults by $\text{MIVS}_{AC}^{1-2}$; these correspond predominantly to stealthy FDIAs that introduce only a small perturbation to $I_2$, resulting in a differential current trajectory that closely resembles that of a high-impedance fault. Similarly, the small percentage of genuine faults misclassified as FDIAs correspond mainly to high-impedance faults at locations close to the relay terminal, where the resulting differential current is small and may resemble a mild FDIA perturbation. These edge cases represent the inherent overlap between the most conservative FDIA profiles and the most subtle fault conditions. The small number of such cases does not compromise the practical utility of the MIVS, as the undetected FDIAs represent only those attacks that produce a differential current signature nearly indistinguishable from a genuine fault, and the misclassified faults can be cleared by backup protection at the cost of a modest increase in fault clearance time. Fig. \ref{fig:samples} illustrates examples of FDIA and fault cases.}
Overall, the results demonstrate that the proposed \Ablack{MIVS} effectively enhances relay security by preventing false trips under FDIA scenarios while maintaining high dependability for internal fault detection in both AC and DC LCDRs.

\begin{table}[t!]
    \centering
    \caption{Confusion Matrices for $MIVS_{AC}^{1-2}$ and $MIVS_{DC}^{34-35}$}
    \renewcommand{\arraystretch}{1.3}

 \textcolor{white}{.......}  $MIVS_{AC}^{1-2}$  \textcolor{white}{..............................................} $MIVS_{DC}^{34-35}$ \\ \textcolor{white}{.......................} 
 
    \begin{minipage}{0.25\textwidth}
        \centering
        \begin{tabular}{ c c c c}
            \Xhline{3\arrayrulewidth}
            &  & \multicolumn{2}{c}{  \textbf{Predicted case}} \\
            \rule{0pt}{3ex}  & & \textbf{Faults} & \textbf{FDIAs} \\
               {\rotatebox{90}{\textbf{case}}}
               &
               {\rotatebox{90}{\textbf{Faults}}}

                & 99.5\% & 0.5\% \\
               {\rotatebox{90}{\textbf{True}}} 
               &
             
               {\rotatebox{90}{\textbf{FDIAs}}}
               
               & 1.7\% & 98.3\% \\
            \Xhline{3\arrayrulewidth}
        \end{tabular}
    \end{minipage}\hfill
    \begin{minipage}{0.2\textwidth}
        \centering
        \begin{tabular}{ c c c c}
            \Xhline{3\arrayrulewidth}
             & & \multicolumn{2}{c}{\textbf{Predicted case}} \\
            \rule{0pt}{3ex} &  & \textbf{Faults} & \textbf{FDIAs} \\
               {\rotatebox{90}{\textbf{case}}}
               
 &
               {\rotatebox{90}{\textbf{Faults}}}
               
               & 100\% & 0\% \\
               {\rotatebox{90}{\textbf{True}}} 
               &
               {\rotatebox{90}{\textbf{FDIAs}}}               
               & 0.07\% & 99.17\% \\
            \Xhline{3\arrayrulewidth}
        \end{tabular}
    \end{minipage}
\label{tab:Confusion_AC_DCR}
\end{table}

\begin{table}[t!]
    \centering
    \caption{Performance Metrics for $MIVS_{AC}^{1-2}$ and $MIVS_{DC}^{34-35}$}
    \renewcommand{\arraystretch}{1.3}
    \begin{tabular}{ c c c c c }
       \Xhline{3\arrayrulewidth}
        \textbf{MIVS} & \textbf{Accuracy} & \textbf{Precision} & \textbf{Recall} & \textbf{F1-Score} \\ \hline 
       
\rule{0pt}{3ex}         \textbf{$MIVS_{AC}^{1-2}$} & 98.9\% & 99.5\% & 98.3\% & 98.9\% \\  
\rule{0pt}{3ex}         \textbf{$MIVS_{DC}^{34-35}$} & 99.6\% & 100\% & 99.2\% & 99.6\% \\  
       \Xhline{3\arrayrulewidth}
    \end{tabular}
    \label{tab:Performance_AC_DC}
\end{table}

\section{Scalability and Sensitivity Analyses}
\label{section:Scalability_Sensitivity}

\subsection{Scalability Analysis: Case Studies 3–41} \label{section:system-wide}

\Ablack{In this subsection, we focus on evaluating the scalability of the proposed \Ablack{MIVS}. Therefore, we consider an alternative system-level approach in which the microgrid security designers/planners train one \Ablack{MIVS} to secure all LCDRs of the same type, i.e., AC or DC, within the inverter-based microgrid, instead of training one \Ablack{MIVS} per LCDR, which can save on training efforts. After training, the \Ablack{MIVS} can be employed in individual LCDRs of the same type, and tested for FDIAs and faults against each LCDR. 
This ensures the \Ablack{MIVS}'s versatility in securing LCDRs of the same type installed anywhere in the same microgrid.
%
%
%
Subsequently, two different \Ablack{MIVS}s are trained: $MIVS_{AC}$ for the AC LCDRs in the inverter-based microgrid, and $MIVS_{DC}$  for the remaining DC LCDRs. 
A comprehensive dataset of fault and FDIA scenarios are simulated for each LCDR in the inverter-based microgrid following the same approach explained in the previous subsection.
In total, 25,344 and 3,456 fault scenarios are simulated for the AC and DC LCDRs, respectively.
Similarly, 25,600 and 3,600 FDIAs are simulated for the AC and DC LCDRs, respectively. }
\Ablack{For each LCDR, fault  scenarios are labeled, shuffled, and split into 80\% for training and 20\% for testing. FDIA scenarios are also labeled and split in the same way.
Afterward, all AC fault and FDIA \textit{training} datasets are combined together in $D_{train}^{AC}$. Similarly, DC fault and FDIA training datasets are combined, forming $D_{train}^{DC}$. 
Testing datasets remain separate, to allow testing for individual LCDRs. 
This yields 32 testing datasets for the AC LCDRs  and 6 testing datasets for the DC LCDRs. 
Following this,   $MIVS_{AC}$  is trained on $D_{train}^{AC}$, while $MIVS_{DC}$, is trained on 
$D_{train}^{DC}$, following the approach described earlier.}

\begin{figure*}[t!]
    \centering
    \includegraphics[width=1.0\textwidth]
    {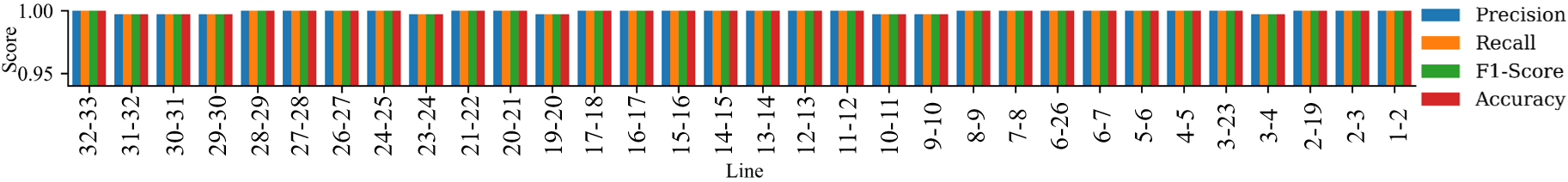}
    \caption{ Performance metrics of the AC-side \Ablack{MIVS}.}
    \label{fig:AC_results}
\end{figure*}

\textit{1) Testing Results of $MIVS_{AC}$:}  
Fig. \ref{fig:AC_results} depicts the results obtained for $MIVS_{AC}$ against FDIAs and faults for all lines in the AC side of the test system. 
They confirm that the proposed \Ablack{MIVS} can accurately detect FDIAs that may target LCDRs installed on any line of the AC side while maintaining the protective dependability, i.e., the fault detection accuracy of the LCDRs across all lines.
Moreover, Fig. \ref{fig:AC_results} reveals that $MIVS_{AC}$ exhibits high performance in terms of all metrics -- precision, recall, accuracy, and $F1$-score.
The minimum observed Accuracy in the AC side is 99.67\%.
We assert that this accuracy stems from the use of RNNs on LCDR current measurement samples within a few milliseconds before and after the triggering event.
The inter-line variation of the results can be explained by the difference in the lines' power flow, i.e., the currents passing through these lines before and after the faults or FDIAs. When a fault occurs, depending on factors such as the fault location, the system contributes, to some extent, to the fault currents, i.e., the local and remote fault currents ($I_1$ and $I_2$) of the LCDR protecting this line.
As a result, some FDIA cases can be confused by the \Ablack{MIVS} with faults. For example, the difference between the local and remote currents of a high-loaded line during a fault may be confused with the difference between the local and remote currents of a lightly-loaded line during an FDIA that slightly increases $I_2$ without affecting $I_1$.
These results reflect the accuracy of the  \Ablack{MIVS} in detecting FDIAs targeting AC LCDRs without affecting their dependability.

\begin{figure}[t!]
\centering
\includegraphics [width=1.0\columnwidth] {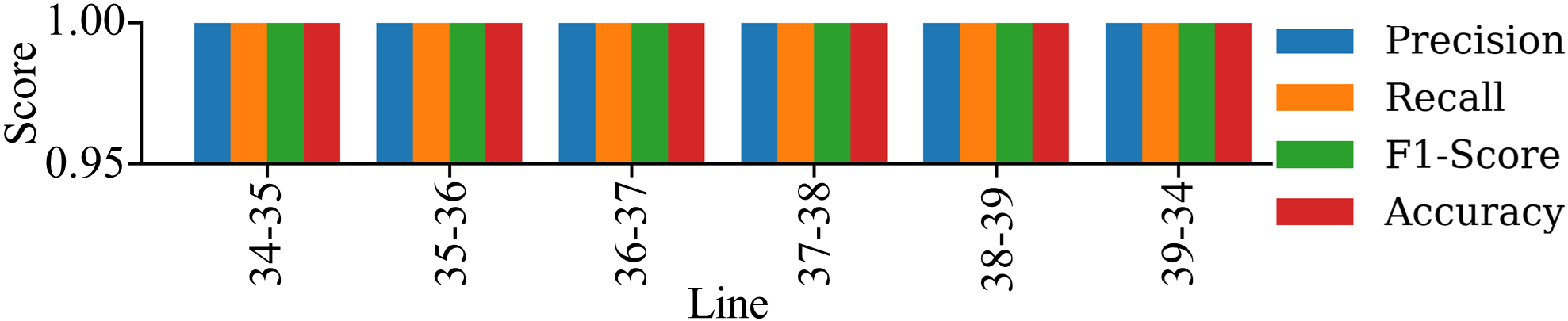}
\caption{ Performance metrics of the DC-side \Ablack{MIVS}.}
  \label{fig:DC_results_A}
\end{figure}


\textit{2) Testing Results of $MIVS_{DC}$:}   
Similarly, the second \Ablack{MIVS}, $MIVS_{DC}$, trained in the previous Section, is tested for each of the 6 lines individually for preciseness. 
Fig. \ref{fig:DC_results_A} summarizes the results obtained in this experiment. 
Our results confirm that the proposed \Ablack{MIVS} can achieve excellent accuracy in detecting more than 99\% of FDIAs and 100\% of faults for all lines in the DC side, underscoring the performance of the \Ablack{MIVS} in securing DC LCDRs against FDIAs.

\subsection{Sensitivity to Measurement Noise: Case Studies 42--82} \label{section:Noise}

\begin{figure*}[t!]
    \centering
    \includegraphics[width=1\textwidth]
    {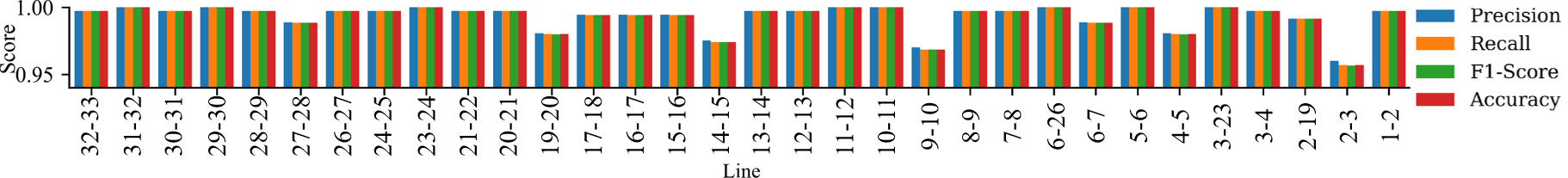} 
    \caption{\Ablue{Performance metrics of the AC-side MIVS considering measurement noise.}}
    \label{fig:AC_results_noise}
\end{figure*}

\begin{figure}[t!]
\centering
\includegraphics [width=1\columnwidth] {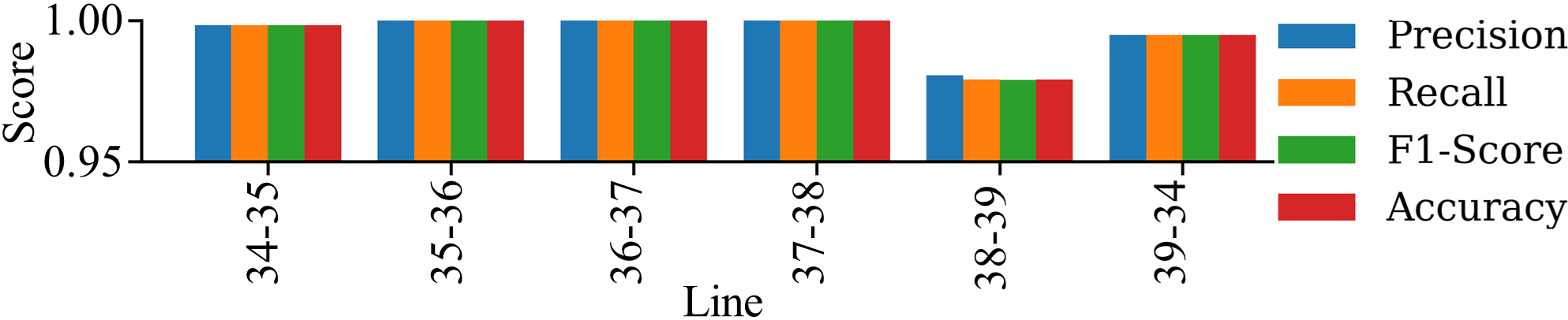}
\caption{ \Ablue{Performance metrics of DC-side MIVS considering measurement noise.}}
  \label{fig:DC_results_noise}
\end{figure}

\Ablue{
Modern LCDRs rely on high-fidelity measurement devices; however, practical measurement systems are inherently subject to noise and acquisition uncertainties. To evaluate the robustness of the proposed MIVS under such conditions, this section investigates its sensitivity to measurement noise ~\cite{noise,gaussian_maths,mypaper_TII}.
Two additional datasets of FDIAs and faults, referred to as noisy datasets, are generated following the same procedure described in the above subsection, and used to evaluate $MIVS_{AC}$ and $MIVS_{DC}$. Measurement noise is modeled as additive white Gaussian noise with a signal-to-noise ratio (SNR) of 40~dB, which represents typical measurement conditions in practical systems~\cite{noise,gaussian_maths,mypaper_TII}. The noise is applied to both local and remote current measurements.
The trained models are evaluated on the noisy datasets without retraining, to assess their inherent robustness to measurement perturbations. As shown in Fig.~\ref{fig:DC_results_noise},  $MIVS_{DC}$ maintains strong performance under noisy conditions, achieving a minimum accuracy of 98.26\% across all tested lines. Similarly, $MIVS_{AC}$ demonstrates robust performance, as illustrated in Fig.~\ref{fig:AC_results_noise}, with a minimum accuracy of 97.13\%.
The observed variations across different scenarios can be attributed to differences in line loading conditions and fault characteristics. In particular, FDIA cases involving minimal perturbations to remote current measurements on heavily loaded lines may produce temporal patterns similar to those of high-impedance faults, leading to occasional misclassifications.
Overall, the results indicate that the proposed MIVS exhibits strong robustness to measurement noise, with only a marginal degradation in performance under realistic noise conditions.}

%
\section{Real-Time Validation and Discussion }
\label{section:hil}


\subsection{Verification  through Real-Time Simulation}\label{section:RTS}

\begin{figure}[t!]
\centering
\includegraphics [width=1.0\columnwidth] {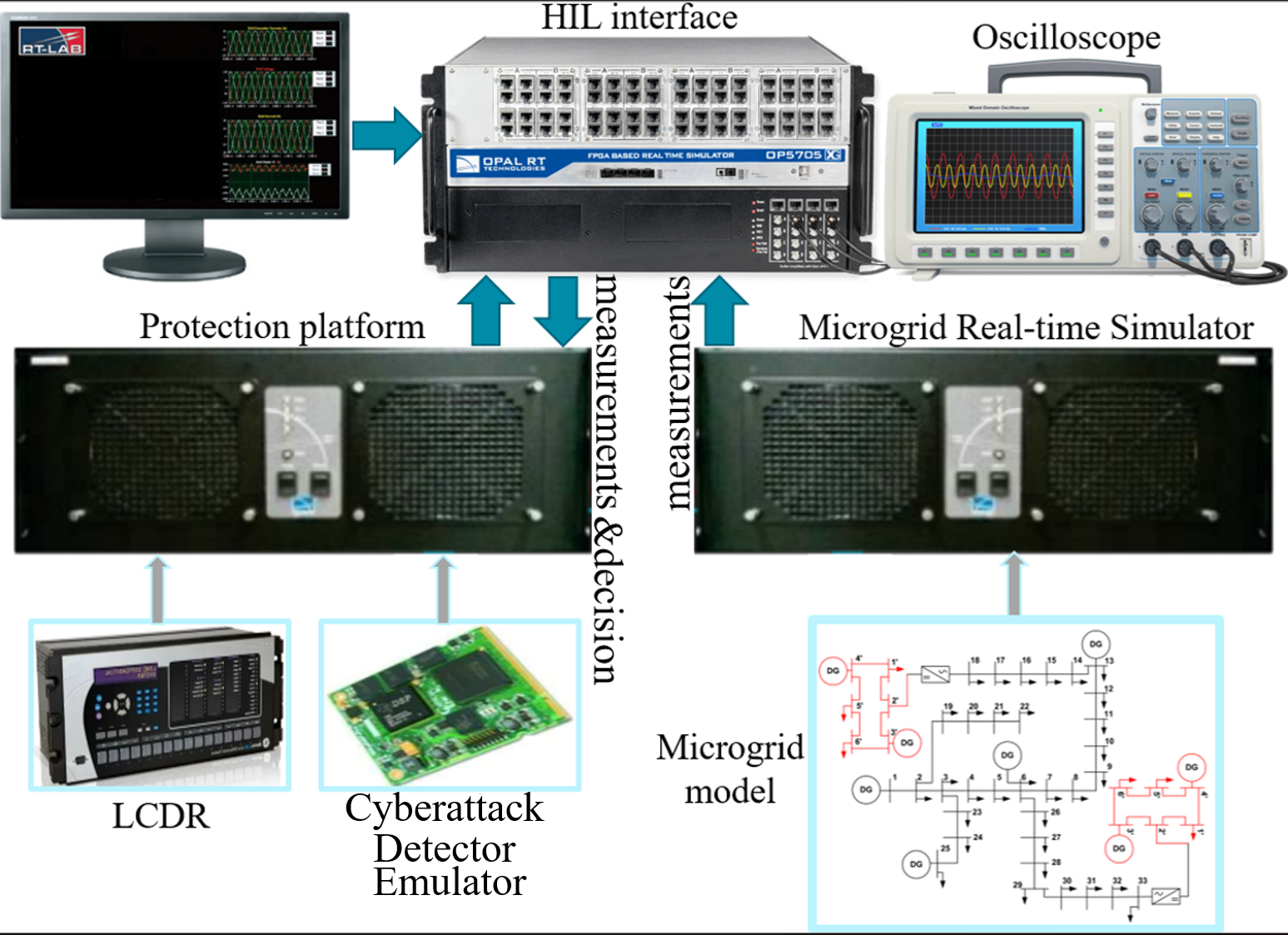}
\caption{Real-Time Simulation Setup. } 
  \label{fig:OPAL_setup}
\end{figure}

\Ablack{To assess the real-time feasibility of the proposed cyber-resilient validation scheme, hardware-in-the-loop (HIL) experiments are conducted using the OPAL-RT real-time simulation platform shown in Fig.~\ref{fig:OPAL_setup}. The setup consists of: (i) an OP5700 real-time digital simulator (RTS) with FPGA-based HIL capability, (ii) a Tektronix DPO4054B digital oscilloscope for high-resolution timing measurements, and (iii) a host PC for model deployment and monitoring.}

\Ablack{The OP5700 RTS integrates reconfigurable FPGA resources and Intel Xeon E5 quad-core processors operating at 2.3–3.0~GHz, enabling deterministic real-time execution of protection algorithms \cite{OP5700}. The trained $CRVS^{1\text{--}2}$ and its associated $LCDR_{1\text{--}2}^{AC}$ are deployed to the RTS using RT-LAB, which automatically converts the developed models into optimized C code suitable for real-time execution.
Within the RTS, the LCDR logic and the CRVS are executed on the same CPU core to reflect a realistic relay-level deployment scenario. A second core is used to simulate the remainder of the microgrid, including inverter-interfaced DGs and network dynamics. The entire system operates synchronously at a sampling frequency of 1~kHz, consistent with practical digital protection implementations.
The CRVS generates a binary validation signal indicating whether the LCDR triggering event corresponds to a genuine internal fault or an FDIA. This signal is routed through the RTS I/O interface and captured by the oscilloscope to accurately quantify the end-to-end detection latency.}

\begin{figure}[t!]
\centering
\includegraphics [width=1\columnwidth] {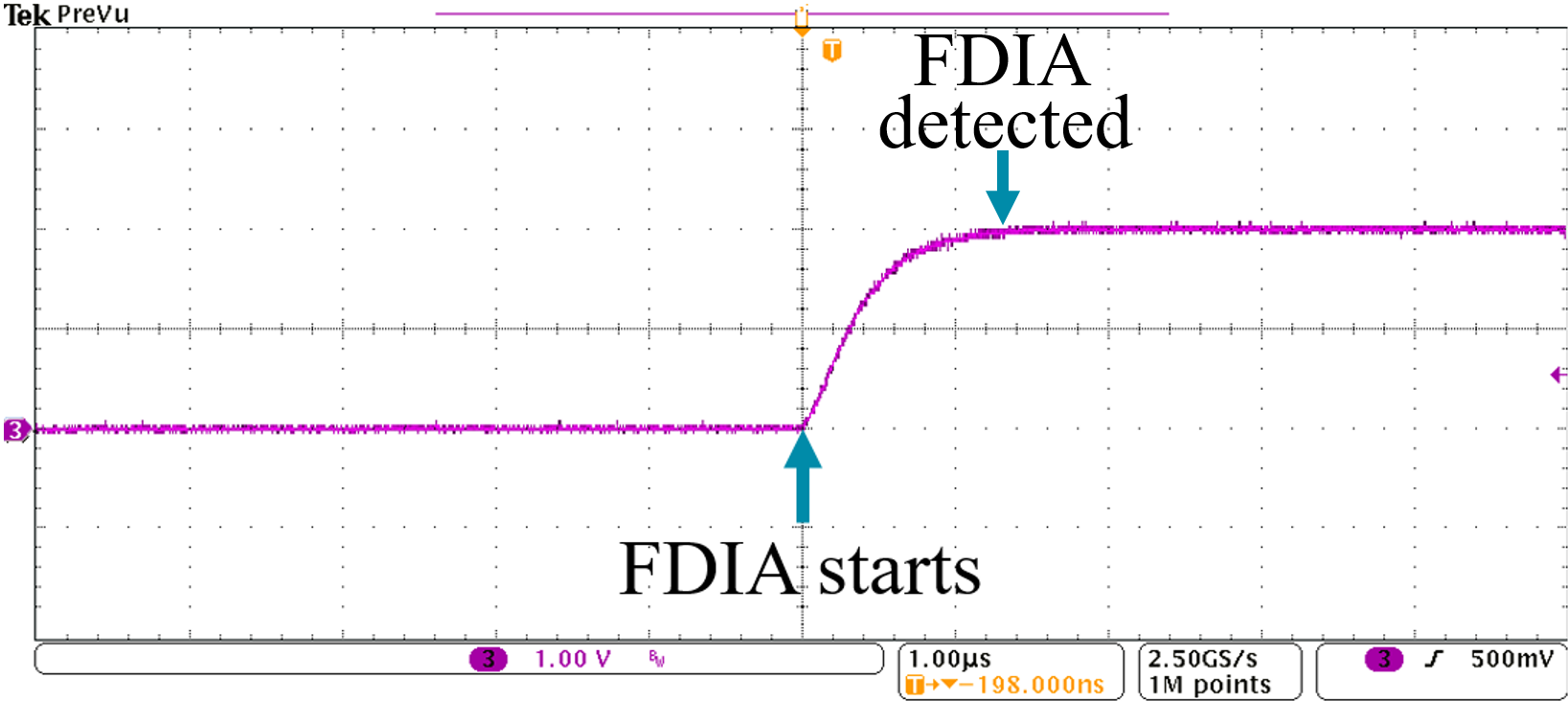}
\caption{Time taken by the proposed \Ablack{MIVS} to detect an FDIA.} 
  \label{fig:OPAL_Oscilloscope}
\end{figure}

\Ablack{Using this setup, multiple fault and FDIA scenarios are executed in real time. Fig.~\ref{fig:OPAL_Oscilloscope} illustrates a representative FDIA case, showing the elapsed time between LCDR pickup and CRVS decision output. The measured detection latency of the CRVS is approximately 1.2~ms.
From a protection perspective, this additional latency is well within acceptable limits. Modern line current differential relays typically operate within a few milliseconds \cite{SEL_inc}, and the CRVS executes in parallel with existing relay logic without altering protection thresholds or operating principles. Consequently, the proposed validation layer does not compromise protection speed or fault clearance requirements.
These HIL results confirm that the proposed CRVS can be executed deterministically in real time using commercially available relay-class hardware. The low computational footprint and sub-cycle detection latency demonstrate the practicality of integrating the CRVS as a supervisory cybersecurity layer in inverter-based microgrids.}


\subsection{Discussion}\label{section:Discussion}

Current protection logic of existing LCDRs is designed to respond to internal fault and cannot distinguish them from FDIAs. In other words, without an \Ablack{MIVS}, 100\% of FDIAs can easily cause both AC and DC LCDRs to unnecessarily issue false trip commands to their circuit breakers.
Therefore, the \Ablack{MIVS} greatly enhances the cybersecurity of such LCDRs, and hence the microgrids security from this angle. 
Despite the great reduction in attack surface after implementing \Ablack{MIVS} as  a new security layer, it is important to recall that achieving perfect security is not easy and requires different and diverse security layers$-$an interesting are of future work.  On another note, after implementing the proposed \Ablack{MIVS}, there is a small percentage of undetected faults. However, this small percentage is not concerning as these faults can be detected by backup relaying logics available in modern LCDRs at the cost of increased fault detection time. 
Moreover, while the proposed \Ablack{MIVS} shows an excellent inference time as demonstrated in Section \ref{section:RTS}, future researchers can investigate and optimize the exact hardware requirements of proposed \Ablack{MIVS} and further study its hardware security. Future work can also focus on incorporating features from the communication layer in the proposed \Ablack{MIVS} to further enhance FDIA detection accuracy.

\begin{table*}[t!]
\renewcommand{\arraystretch}{1.3}

    \centering
        \caption{Qualitative Comparison with Related Works Focusing on Microgrid LCDRs}
    \begin{tabular}{c c c c c} \Xhline{3\arrayrulewidth}
\rule{0pt}{3ex}  \textbf{Point of Comparison}  & \textbf{Proposed Approach} & \textbf{\cite{Amir4}} & \textbf{\cite{ResMVDC}} & \textbf{\cite{Cyber_Resilient_Protection}} \\ 
       \hline

\rule{0pt}{3ex}   
         - Model-free approach?  & Yes& No& No& No \\ 
 
\rule{0pt}{3ex} 
         - Applicability &  AC \& DC LCDRs   &  DC  LCDRs&  DC  LCDRs  &  DC LCDRs     \\   

\rule{0pt}{3ex}          - Required additional microgrid components  & None  &  Extra  reactors \&capacitors   & Extra reactors  & LCDR's line  \\ 
         or special configurations  &&  for each protected line  & for each protected line& must be bipolar  \\  
         
\rule{0pt}{3ex} 
- Speed of FDIA-detection & $<$ 2ms & not discussed & not discussed & up to 10 ms  \\ 
        \Xhline{3\arrayrulewidth}
    \end{tabular}
    \label{table:comparative_analysis} 
\end{table*}




                        


\begin{table}[t!]
    \centering

    \caption{ Quantitative Comparison with Related Works Proposing Machine Learning-Based Solutions for LCDRs }

    \renewcommand{\arraystretch}{1.3}

    \setlength{\tabcolsep}{4pt} 
    \begin{tabular}{c c c c} 
        \Xhline{3\arrayrulewidth}
        \rule{0pt}{3ex}  \textbf{Case} & \textbf{Proposed \Ablack{MIVS}} &  \cite{mypaper_TII,Amir_ML}'s \textbf{approach} &  \cite{Ahmad_smartgrid}'s \textbf{approach} \\ 
        \hline
        \rule{0pt}{3ex}         
        $LCDR_{1-2}^{AC}$ & \textbf{98.9\%} & 92.79\% & 89.47\%  \\ 
        \rule{0pt}{3ex}         
        $LCDR_{1-2}^{DC}$ & \textbf{99.6\%} & 96.4\% & 91.64\%  \\ 
        \Xhline{3\arrayrulewidth}
    \end{tabular}
    \label{table:perfromance_comparison}
\end{table}

\section{\Ablack{Related Work and Comparative Analysis}}
\label{section:related}

\Ablack{Related works have explored the use of AI and deep learning for measurement interpretation in power systems, including fault detection in transformers, wide-area damping controllers, dynamic state estimation, and transmission line protection. For instance, Thomas \textit{et al.} \cite{Thomas_TIM_2023} proposed a CNN-transformer model to extract features from time-domain current measurements for high-impedance fault detection in distribution networks. Asghari \textit{et al.} \cite{CyberResilient_TIM_DistanceRelay} introduced a cyber-resilient random-forest-based scheme for transmission line protection using traveling-wave measurements, highlighting the potential of AI to assess the physical consistency of measured signals. Similarly, Saber \textit{et al.} \cite{saber2025model} demonstrated a data-driven approach to detect adversarial attacks on transformer differential relays, further illustrating how AI can act as a supervisory measurement validation layer. Other TIM studies, such as \cite{ali2023reliable,zadsar2023preventing,riahinia2024adaptive}, have applied ensemble learning and adaptive penalized methods to identify anomalies in transformer and wide-area measurement streams under adversarial conditions.}

\Ablue{The methods in~\cite{Amir4} and~\cite{ResMVDC} require the installation of extra passive components---reactors and capacitors---on each protected line, which increases hardware cost and is impractical for large-scale microgrid deployments. Both methods are also limited to DC LCDRs and cannot be extended to AC line protection. The approach in~\cite{Cyber_Resilient_Protection} avoids additional hardware but requires the protected DC line to be bipolar, restricting its applicability, and exhibits a detection time of up to 10\,ms, which may be insufficient for fast-acting protection systems. Machine-learning-based methods such as those in~\cite{mypaper_TII,Amir_ML} rely on features derived from current magnitudes and phase angles, which are effective in transmission systems with large fault currents but are less discriminative in inverter-based microgrids where fault current magnitudes are actively capped. Similarly, the principal-component-based isolation-forest approach of~\cite{Ahmad_smartgrid} is trained on steady-state current magnitude snapshots, making it less sensitive to the temporal structure of transient fault waveforms characteristic of inverter-dominated systems. In contrast, the proposed MIVS operates directly on short windows of instantaneous current measurements, requires no additional hardware, applies uniformly to both AC and DC LCDRs, and achieves sub-2\,ms detection latency, without requiring hand-crafted features or system-topology-specific configurations.}
\Ablue{While these works illustrate the general applicability of AI to measurement validation in the power domain, none specifically address the challenges of LCDRs in \textit{inverter-based microgrids}.} In particular, inverter-based microgrid LCDRs operate with low fault currents due to inverter-imposed limits, making traditional magnitude-based or phasor-based detection techniques ineffective. The proposed \Ablack{MIVS} is, to the best of our knowledge, the first AI-based measurement validation framework designed to interpret short-window synchronized multi-phase current measurements from both AC and DC LCDRs in inverter-based microgrids, distinguishing physically plausible fault-induced currents from cyber-manipulated streams while preserving relay dependability and real-time operability. This positions the \Ablack{MIVS} as a novel extension of the TIM literature on AI for measurement and instrumentation, expanding its application to microgrid protection instrumentation under adversarial conditions.

\Ablack{Moreover,  we compare the proposed \Ablack{MIVS} with the most related works on LCDRs. 
Compared to previous works on  LCDRs installed in microgrid in general, i.e., \cite{Cyber_Resilient_Protection,Amir4, ResMVDC}, the proposed approach is the only approach that also does not require additional microgrid components and maintains 
an acceptable FDIA detection time.
In further detail,} as explained in Section \ref{section:Introduction}, one of the main advantages of the proposed method is that it does not require the installation of additional microgrid components, unlike existing techniques, e.g., \cite{Amir4, ResMVDC}, which are also specific to DC microgrids only \Ablack{\textit{and cannot be applied to detect FDIAs on AC LCDRs.}}
Another advantage of the proposed method is its high speed, as it can detect FDIAs and confirm faults in less than 2 milliseconds, as verified by the real-time experiment in Section \ref{section:RTS}. Therefore, the proposed scheme is faster than methods that may increase fault-detection time, such as that in  \cite{Cyber_Resilient_Protection}. 
The method in \cite{Cyber_Resilient_Protection} \textit{cannot be applied to detect FDIAs on AC LCDRs,} applies only to DC microgrids and only when the protected line is bipolar. 
A summary of this analysis is depicted in Table     \ref{table:comparative_analysis}.

Additionally, this section quantitatively compares the performance of the \Ablack{MIVS} with existing machine-learning-based approaches for LCDRs in transmission systems, as no studies specifically address LCDRs in inverter-based microgrids. Two primary methods are considered: 1) MLPs trained on features derived from current magnitudes and phase angles \cite{mypaper_TII,Amir_ML}, and 2) principal component analysis with an isolation-forest algorithm trained on current magnitudes \cite{Ahmad_smartgrid}.
It is worth-noting to recall that transmission system LCDRs experience significantly higher fault currents that affect the LCDR's local and remote current measurements compared to FDIAs, making it easier to distinguish faults from cyberattacks using current magnitude variations \cite{Ahmad_smartgrid}. This assumption, however, does not hold in inverter-based microgrids, where fault currents are limited and fault level is much lower than that of transmission systems. The proposed \Ablack{MIVS} is compared with these methods by applying them to $LCDR_{1-2}^{AC}$ and $LCDR_{34-35}^{DC}$. The results, shown in Table \ref{table:perfromance_comparison}, confirm that the proposed \Ablack{MIVS} outperforms existing approaches in FDIA detection accuracy while maintaining LCDR dependability and speed. This improvement is attributed to the \Ablack{MIVS}’s ability to learn patterns in both local and remote currents during faults and FDIAs, rather than relying solely on static snapshots of current magnitudes (or angles), as in previous methods.

\section{Conclusion} \label{section:Conclusion}

This paper proposed a novel and flexible \Ablack{MIVS} that can be used to detect FDIAs against AC and DC LCDRs in inverter-based microgrids. 
The proposed \Ablack{MIVS} requires only LCDR current measurements. 
After implementation, the LCDR trips only if the \Ablack{MIVS} confirms that the triggering event is an actual fault, mitigating the impact of FDIAs.
The proposed \Ablack{MIVS} leverages an RNN that exploits the time dependence of LCDR current measurements instead of relying on hand-crafted features unsuitable for inverter-based microgrids.
%
The performance of the proposed \Ablack{MIVS} is evaluated using an inverter-based microgrid test system
under an extensive set of faults and FDIAs.
%
Our results indicate that the \Ablack{MIVS}: 1) can accurately detect FDIAs on both AC and DC LCDRs, outperforming existing methods, 2) does not greatly affect the LCDR's dependability, and 3)
is robust to system variations.
%
The \Ablack{MIVS}'s ability to operate in real-time was verified using OPAL-RT's simulator. Future work directions have also been discussed.


\bibliographystyle{IEEEtran} 
\bibliography{_References}

\begin{IEEEbiography} [{\includegraphics[width=1in,height=1.25in,clip,keepaspectratio]{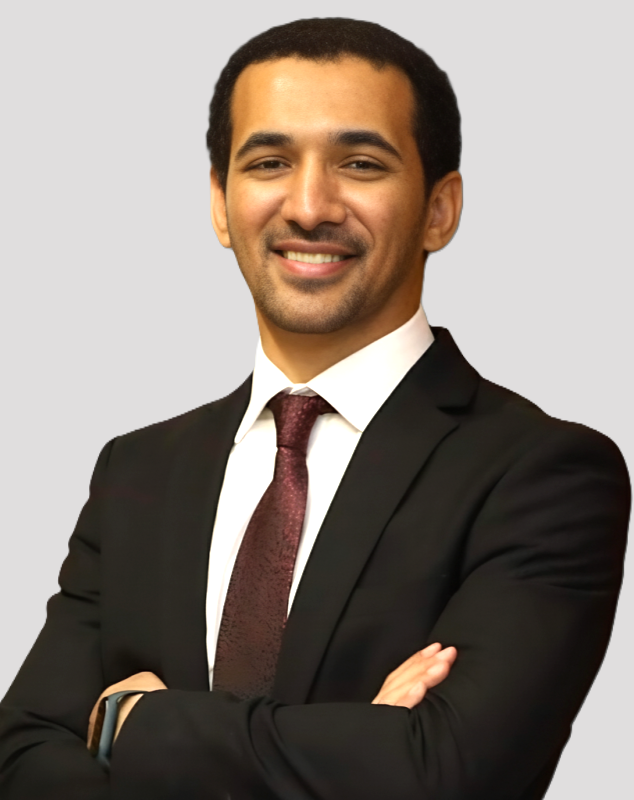}}]{Ahmad Mohammad Saber}
(Member, IEEE) received the B.Sc. degree from Ain Shams University, Egypt, in 2016, the M.Sc. degree from Cairo University, Egypt, in 2019, and the Ph.D degree from Khalifa University, UAE, in 2024. 
He is currently a Postdoctoral fellow at the University of Toronto, ON, Canada. 

He received the Outstanding Thesis Award at IEEE SmartGridComm 2025, and the UAE Excellence and Creative Engineering Award in 2024.
Since 2016, he has held technical and commercial roles in several industries including power transformers manufacturing, water and wastewater treatment, and access control and security systems. In 2023, he was an international visiting graduate student at the University of Toronto, ON, Canada.  His current research interests include cyber-physical security, AI applications and security, power system protection, distributed generation, and renewable power planning and integration. 
Dr. Saber is a reviewer for multiple IEEE Transactions journals and a technical program committee member for the IEEE PST 2026.
\end{IEEEbiography}

\begin{IEEEbiography} [{\includegraphics[width=1in,height=1.25in,clip,keepaspectratio]{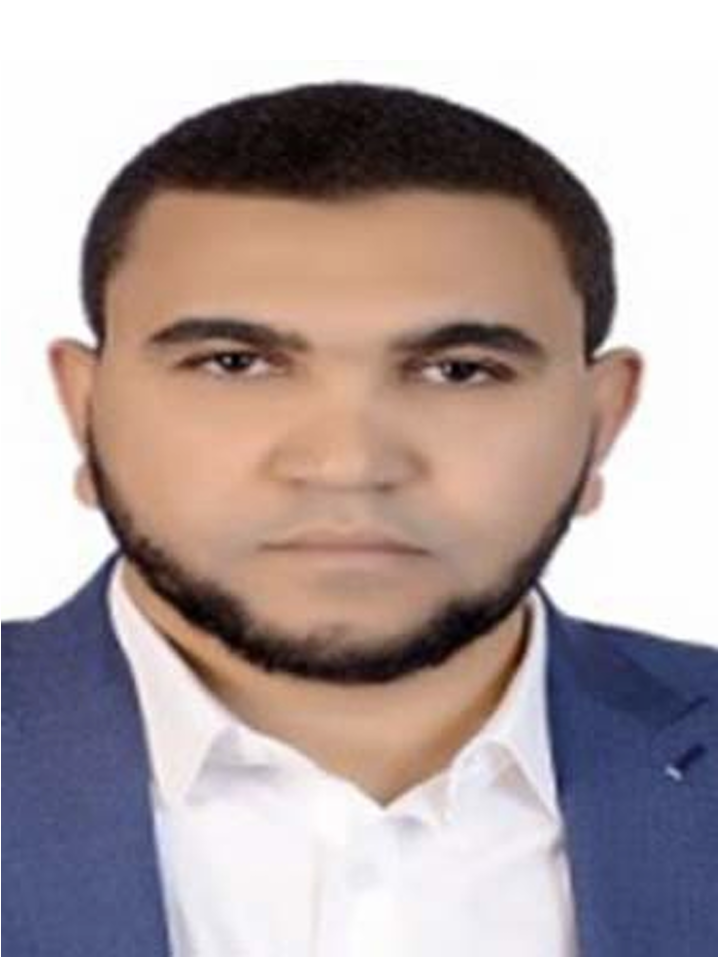}}]{Ahmed Saber Refae}
received the B.Sc., M.Sc., and Ph.D. degrees in electrical power engineering from Cairo University, Giza, Egypt, in 2008, 2013, and 2019, respectively.He was a Postdoctoral Fellow with Khalifa University, Abu Dhabi, UAE. He is currently an Associate Professor with Electrical Power Engineering Department, Faculty of Engineering, Cairo University. His research interests include HVAC and HVDC transmission system protection, AC microgrid protection, and hybrid AC/DC microgrid protection.
\end{IEEEbiography}

\begin{IEEEbiography} [{\includegraphics[width=1in,height=1.25in,clip,keepaspectratio]{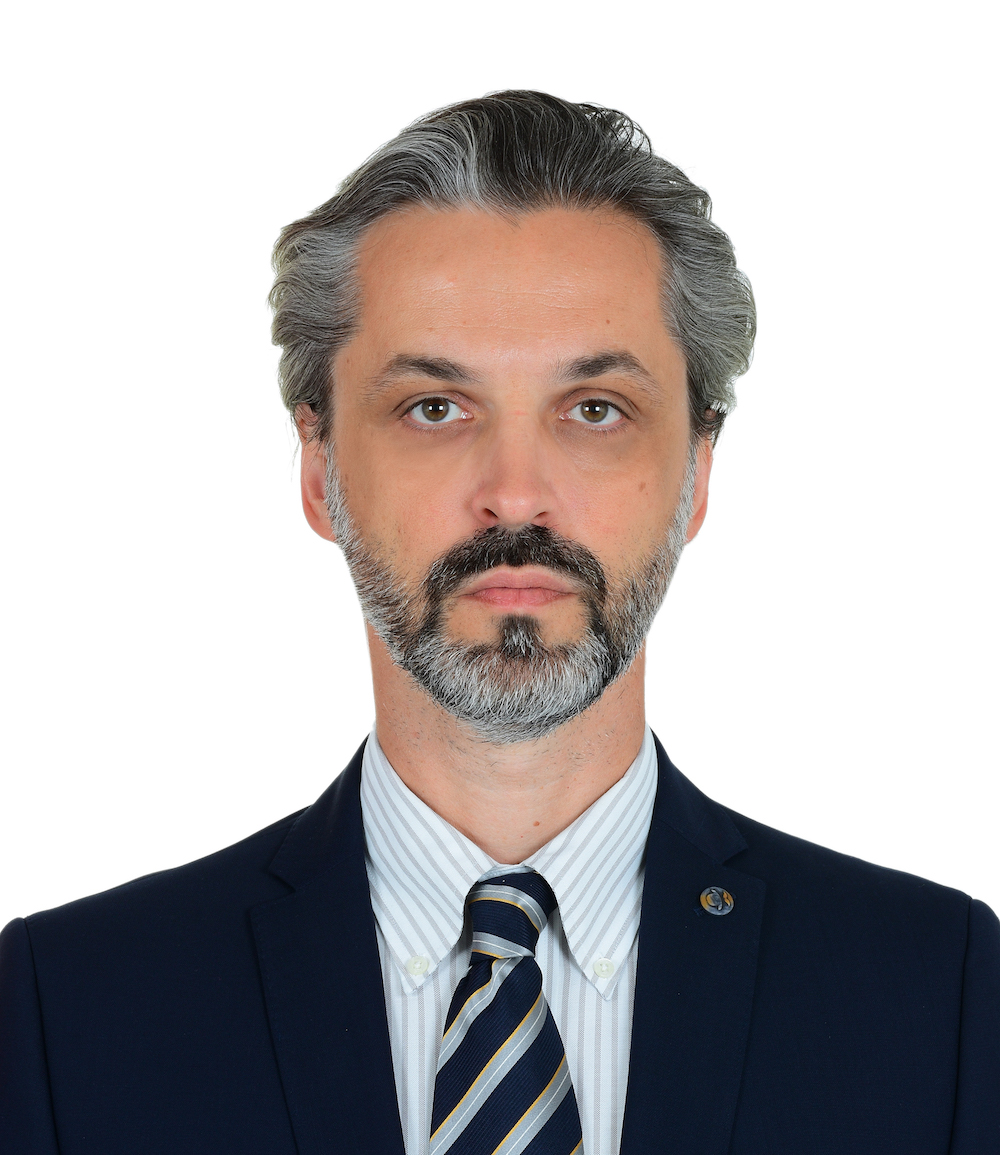}}]{Davor Svetinovic}
(SM’16) is a professor of computer science at the Department of Computer Science, Khalifa University, Abu Dhabi, and a visiting fellow at the ADIA Lab, Abu Dhabi, UAE. He received his doctorate in computer science from the University of Waterloo, Waterloo, ON, Canada, in 2006. Previously, he worked at WU Wien, Austria, TU Wien, Austria, and Lero $-$ the Irish Software Engineering Center, Ireland. He was the head and director of the Research Center for Cryptoeconomics in Vienna, Austria. He was a visiting professor and a research affiliate at MIT and MIT Media Lab, MIT, USA. Davor has extensive experience working on complex multidisciplinary research projects. He has published more than 120 papers in leading journals and conferences and is a highly cited researcher in blockchain technology. His research interests include cybersecurity, blockchain technology, cryptoeconomics, trust, and software engineering. His career has furthered his interest and expertise in developing advanced research capabilities and institutions in emerging economies. He is a Senior Member of IEEE and ACM (Lifetime) and a Mohammed Bin Rashid Academy of Scientists affiliate.
\end{IEEEbiography}
\begin{IEEEbiography} [{\includegraphics[width=1in,height=1.25in,clip,keepaspectratio]{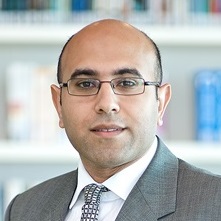}}]{Hatem H. Zeineldin}
 (M’06–SM’13) received the B.Sc. and M.Sc. degrees in electrical engineering from Cairo University, Giza, Egypt, in 1999 and 2002, respectively, and the Ph.D. degree in electrical and computer engineering from the University of Waterloo, Waterloo, ON, Canada, in 2006. He was with Smith and Andersen Electrical Engineering, Inc., North York, ON, USA, where he was involved in projects involving distribution system designs, protection, and distributed generation. He was a Visiting Professor with the Massachusetts Institute of Technology, Cambridge, MA, USA. He is with Khalifa University of Science and Technology, Abu Dhabi, UAE and on leave from Faculty of Engineering, Cairo University. His current research interests include distribution system protection, distributed generation, and microgrids. 
\end{IEEEbiography}

\begin{IEEEbiography} [{\includegraphics[width=1in,height=1.25in,clip,keepaspectratio]{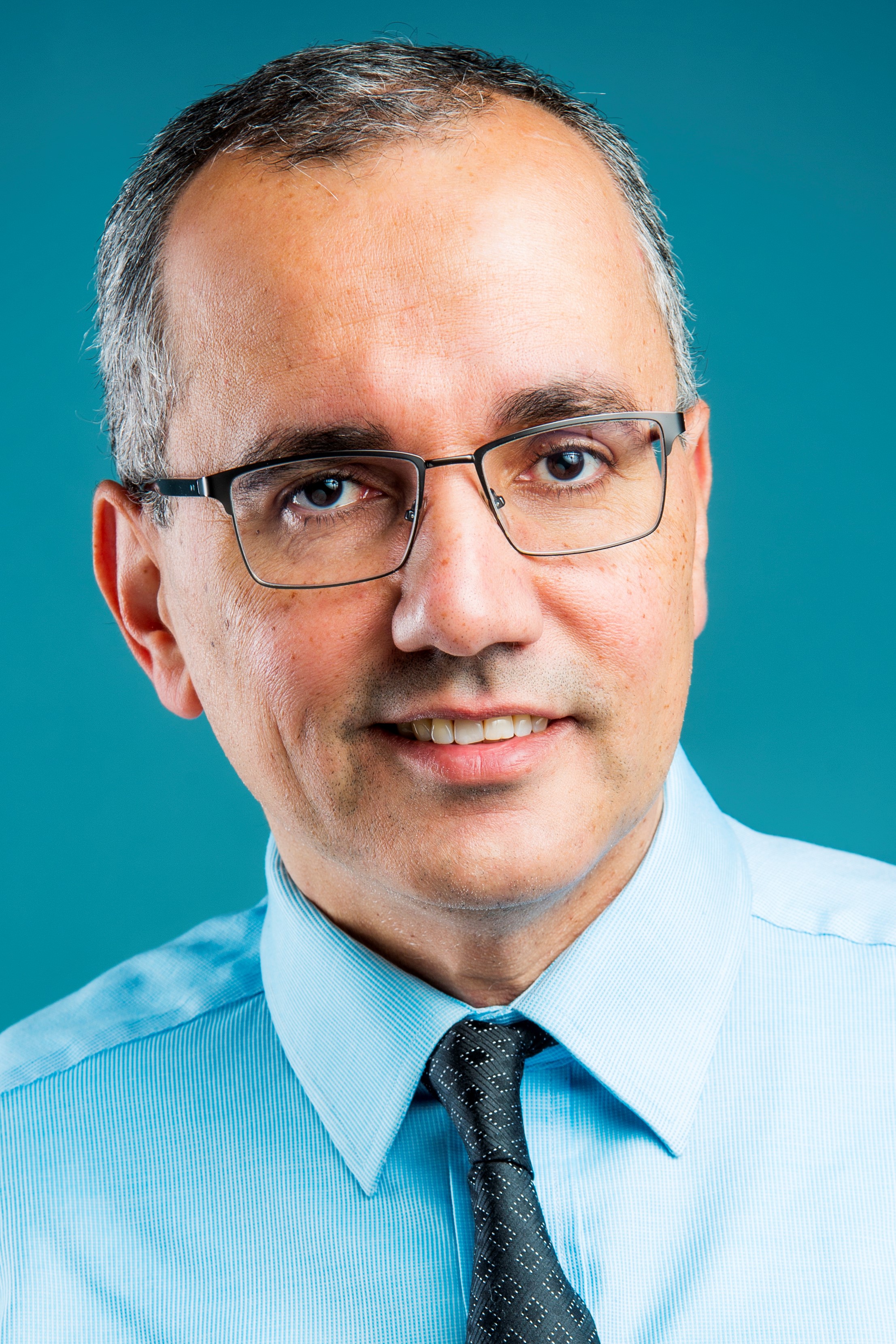}}]{Amr Youssef}
(SM’06) received the B.Sc. and M.Sc. degrees from Cairo University, Cairo, Egypt, in 1990 and 1993, respectively, 
and the Ph.D. degree from Queens University, Kingston, ON, Canada, in 1997. He was with Nortel Networks, the Center
for Applied Cryptographic Research, University of
Waterloo, IBM, and also with Cairo University. He is
currently a Professor with the Concordia Institute for
Information Systems Engineering, Concordia University, Montreal, Canada. 
He has authored over 300 referred journal and conference publications in areas related to his research interests. 
His current research interests include information security, and cyber-physical systems security. 
Dr. Youssef  served on over 100 technical program committees of cryptography and data security conferences. 
He was the co/chair for Africacrypt 2010, Africacrypt 2020, the conference Selected Areas in Cryptography (SAC 2014, SAC 2006, and SAC 2001).
\end{IEEEbiography}

\begin{IEEEbiography} [{\includegraphics[width=1in,height=1.25in,clip,keepaspectratio]{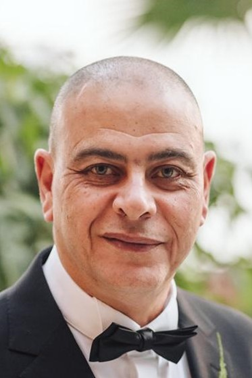}}]{Ehab El-Saadany}
 (F’18) is an IEEE Fellow for his contributions in distributed generation planning, operation and control. He received his B.Sc. and M.Sc. degrees in Electrical Engineering from Ain Shams University, Cairo, Egypt, in 1986 and 1990, respectively, and his Ph.D. degree in Electrical Engineering from the University of Waterloo, Waterloo, ON, Canada, in 1998, where he was a Professor with the ECE Department till 2019, where he was the Director of the Power MEng program between 2010 and 2015. Currently, he is a Professor at the Department of Electrical Engineering and the Dean of College of Engineering and Physical Sciences at Khalifa University, UAE, and an and Adjunct Professor at the ECE Department, University of Waterloo, Canada.  Dr. El-Saadany is an internationally recognized expert in the area of sustainable energy integration and smart distribution systems. His research interests include smart grid operation and control, microgrids, transportation electrification, self-healing, cyber-physical security of smart grids, protection, power quality, and embedded generation. He is an Editor of the IEEE TRANSACTIONS ON SMART GRID, the IEEE TRANSACTIONS ON POWER SYSTEMS, and IEEE Power Engineering Letters.  He is a Registered Professional Engineer in the Province of Ontario.
\end{IEEEbiography}

\begin{IEEEbiography} [{\includegraphics[width=1in,height=1.25in,clip,keepaspectratio]{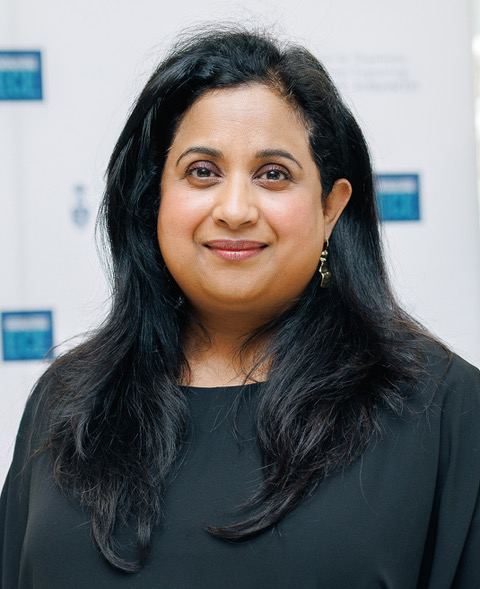}}]{Deepa Kundur}
(Fellow, IEEE) is Professor and Chair of The Edward S. Rogers Sr. Department of Electrical and Computer Engineering (ECE) at the University of Toronto and holds a Tier 1 Canada Research Chair in Cybersecurity of Intelligent Critical Infrastructure. A native of Toronto, Canada, she received her B.A.Sc., M.A.Sc., and Ph.D. degrees in Electrical and Computer Engineering from the University of Toronto in 1993, 1995, and 1999, respectively.

Professor Kundur’s research focuses on the cybersecurity of critical infrastructure, with emphasis on energy and transportation systems, as well as data-centric approaches in psychiatry. Her work spans the modeling and analysis of cyber-physical attacks, advanced detection and inference methods using diverse and multimodal data sources, and cyber-physical strategies for infrastructure resilience. Her research applies methods drawn from deep learning, dynamical systems, applied cryptography, network theory, and optimization. She has authored over 275 publications and has received numerous paper awards.

Her professional service includes roles in national research funding and technical communities. She served as a member and Chair of the NSERC Discovery Grant Evaluation Group in Electrical and Computer Engineering (2013–2020), a primary source of federal research funding in Canada, and has contributed to advisory committees related to cybersecurity of critical infrastructure, as well as conference leadership and editorial roles at major IEEE and ACM venues. She is a Fellow of the IEEE, a Fellow of the Canadian Academy of Engineering, a Fellow of the Engineering Institute of Canada, and a Senior Fellow of Massey College.
\end{IEEEbiography}

\end{document}